\documentclass[twocolumn]{aastex62}
\usepackage[T1]{fontenc}
\usepackage[utf8]{inputenc}
\usepackage{subfiles}
\usepackage{amsmath}
\usepackage{wasysym}

\usepackage{savesym}
\savesymbol{tablenum}
\usepackage[separate-uncertainty=true, detect-weight=true]{siunitx}
\restoresymbol{SIX}{tablenum}

\DeclareSIUnit\parsec{pc}

\turnoffeditone
\turnoffedittwo

\AtBeginDocument{%
}

\AtEndDocument{%
        \bibliography{hubbleIC63}
}

\accepted{November 6, 2019}
\submitjournal{ApJ}

\shortauthors{Van De Putte et al.}

\begin{document}

\title{Evidence of dust grain evolution from extinction mapping in the IC\,63 photodissociation region\footnote{Based on observations made with the NASA/ESA Hubble Space Telescope, obtained at the Space Telescope Science Institute, which is operated by the Association of Universities for Research in Astronomy, Inc., under NASA contract NAS5-26555. These observations are associated with program GO-14186.
}}

\author[0000-0002-5895-8268]{Dries Van De Putte} \affiliation{Sterrenkundig
  Observatorium, Universiteit Gent, Krijgslaan 281 S9, 9000 Gent, Belgium}

\author[0000-0001-5340-6774]{Karl D. Gordon} \affiliation{Space Telescope Science Institute, 3700 San
  Martin Drive, Baltimore, MD21218, USA}
\affiliation{Sterrenkundig
  Observatorium, Universiteit Gent, Krijgslaan 281 S9, 9000 Gent, Belgium}

\author{Julia Roman-Duval} \affiliation{Space Telescope Science Institute, 3700
  San Martin Drive, Baltimore, MD21218, USA}

\author{Benjamin F. Williams} \affiliation{Department of Astronomy, Box 351580,
  University of Washington, Seattle, WA 98195, USA}

\author{Maarten Baes} \affiliation{Sterrenkundig Observatorium, Universiteit
  Gent, Krijgslaan 281 S9, 9000 Gent, Belgium}

\author{Kirill Tchernyshyov} \affiliation{Department of Physics and Astronomy, Johns Hopkins University, 3400 N. Charles Street, Baltimore, MD 21218, USA}

\author{Brandon L. Lawton} \affiliation{Space Telescope Science Institute, 3700
  San Martin Drive, Baltimore, MD21218, USA}

\author{Heddy Arab} \affiliation{Observatoire astronomique de Strasbourg,
  Université de Strasbourg, 11 rue de l’Université, 67000 Strasbourg, France}

\correspondingauthor{Dries Van De Putte}
\email{drvdputt.vandeputte@ugent.be}

\begin{abstract}
  Photodissociation regions (PDRs) are parts of the ISM consisting of
  predominantly neutral gas, located at the interface between H \textsc{ii} regions and
  molecular clouds. The physical conditions within these regions show variations
  on very short spatial scales, and therefore PDRs constitute ideal laboratories for
  investigating the properties and evolution of dust grains. We have mapped
  IC\,63 at high resolution from the UV to the NIR \edit1{(\SI{275}{\nano\metre} to \SI{1.6}{\micro\metre})}, using the Hubble Space Telescope WFC3. Using a Bayesian SED fitting tool, we simultaneously derive a set of stellar ($T_\text{eff}$, $\log(g)$, distance) and extinction ($A_V$, $R_V$) parameters for 520 background stars.
  We present maps of $A_V$ and $R_V$ with a resolution of 25 arcsec based on these results.
  The extinction properties vary across the PDR, with values for $A_V$ between 0.5 and 1.4 mag, and a decreasing trend in $R_V$, going from 3.7 at the front of the nebula to values as low as 2.5 further in. This provides evidence for evolution of the dust optical properties.
  We fit two modified blackbodies to the \edit1{MIR and} FIR SED, obtained by combining the $A_V$ map with data from \textit{Spitzer} and \textit{Herschel}.
  We derive \edit1{effective} temperatures (30 K and 227 K) and the ratio of opacities at \SI{160}{\micro\metre} to V band $\kappa_{160} / \kappa_V$ (\num{7.0e-4} and \num{2.9 e-9}) for the two dust populations.
  Similar fits to individual pixels show spatial variations of $\kappa_{160} / \kappa_{V}$.
  The analysis of our HST data, combined with these \textit{Spitzer} and \textit{Herschel} data, provides the first \edit1{panchromatic} view of dust within a PDR.
\end{abstract}

\section{Introduction}
Photodissociation regions (PDRs) are regions of the interstellar medium (ISM) where the physical
properties of the gas are mainly determined by the radiation field of nearby O or B stars. They
form a boundary layer between ionized (H\textsc{ii}) regions, and the rest of the molecular cloud where
they reside.
The FUV radiation from
ionizing stars is quickly attenuated by the opacity of the relatively high density medium shielding the rest of the gas
\edit1{(several 10s up to $\sim10^6$\,\si{\per\cubic\cm})}.
As a consequence, there are typically sharp transitions between different regimes (e.g.~H\textsc{ii}, H\textsc{i}
and H\textsubscript{2}), resulting in a layered structure \citep{1997ARA&A..35..179H, 1999RvMP...71..173H}. This makes PDRs the perfect
laboratory for studying the evolution of the ISM, including the dust over different physical and excitation conditions.

Interstellar dust grains make a very significant contribution to the total opacity of the
interstellar medium, and modify any \edit1{impinging} radiation field through the effects of absorption,
scattering and re-emission \citep{2003ARA&A..41..241D, 2013ARA&A..51...63S}. Aside from attenuating the UV radiation
field that regulates the physics inside a PDR, they also couple to the gas through other
processes. Dust grains play a major role in the formation of H\textsubscript{2}, through
reactions that take place on grain surfaces \citep{1971ApJ...163..155H, 2017MolAs...9....1W}.
They provide a major contribution to the heating of the gas through the photoelectric effect,
but can also have a cooling effect when gas-grain collision occur \citep{1994ApJ...427..822B,
  2001ApJS..134..263W}.

To model these effects, knowledge about the composition and size distribution of the dust grains
is necessary. While there are many dust models that can explain the observed extinction curves
and emission spectra \edit1{\citep{1990A&A...237..215D, 2004ApJS..152..211Z, 2007ApJ...657..810D, 2011A&A...525A.103C, 2017A&A...602A..46J}}, it remains difficult to accurately constrain the exact properties of the
grains.
Moreover, the effects of dust evolution processes can change these properties depending on the time and the environment.
Some models, such as THEMIS \citep{2017A&A...602A..46J}, have
built-in ways to follow the changes in dust properties. Providing better constraints on these models is crucial for understanding not only the dust itself, but also the structure and
evolution of PDRs and the ISM in general.

Evidence for dust evolution in PDRs has been found through observations in the mid and
far-infrared.
\edit1{By applying Blind Signal Separation methods to \textit{Spitzer} IRS data of several PDRs (Ced 201, NGC 7023 East and North-West, $\rho$ Oph), \citet{2007A&A...469..575B} identified two spectral shapes.
  One mainly contains the Aromatic Infrared Bands (AIBs), and is linked to Polycyclic Aromatic Hydrocarbons (PAHs).
  The other exhibits a combination of broad AIBs and MIR continuum emission, and was found to correspond to Very Small Grain (VGSs).
  Later work by \citet{2014ApJ...795..110B}, uses $k$-means spectral clustering to identify zones in NGC 7023 that have similar spectral shapes, and find a spatial evolution in the PAH band strength ratios.}

  \edit1{Radiative transfer modeling by \citet{2008A&A...491..797C} shows that differences in excitation conditions are not enough to explain the observed variations of the AIBs and MIR continuum in NGC 2023 and the Horsehead nebula, compared to the diffuse ISM.
  Changes in the relative abundances of the PAHs and VSGs are suspected.
In the Orion bar, a radiative transfer model which uses the dust \edit2{abundances} of the diffuse ISM is sufficient to explain the dust variations derived from \textit{Herschel}/PACS/SPIRE and \textit{Spitzer}/IRAC maps, but overestimates the PAH emission at 3.6 \si{\micro\metre} \citep{2012A&A...541A..19A}.
Using theoretical dust emission models which include the coagulation of grains, \citet{2015A&A...579A..15K} were able reproduce the changes in temperature, spectral index, opacity, and MIR emission, that are observed when transitioning from the diffuse ISM to high-density regions.}

\edit1{IC\,63 is a nearby nebula which is illuminated by the B\,IV star $\gamma$~Cas.
  Based on the Hipparcos parallax \citep{1997A&A...323L..49P} of 5.32 mas, the distance to $\gamma$~Cas is roughly 190 pc.}
  A comprehensive study of the physics and chemistry of IC\,63 can be found in
  \citet{1994A&A...282..605J, 1995A&A...302..223J, 1996A&A...309..899J}.
  A study using \textit{Infrared Space Observatory} (ISO) data
 has characterized the fine structure and H\textsubscript{2} lines in IC\,63 \citep{2009MNRAS.400..622T}, and the distribution of Polycyclic Aromatic Hydrocarbons (PAHs) has been derived from \textit{Spitzer} data \citep{2010ApJ...725..159F}. Recently, \textit{Herschel}
FIR maps and FIR spectroscopy data were combined with [C \textsc{ii}] 157 \si{\micro\meter} velocity maps
from the GREAT instrument on board the SOFIA observatory to revisit IC\,63 in greater detail
\citep{2018A&A...619A.170A}.

\edit1{For the radiation field at the tip of IC\,63, the model of \citet{1995A&A...302..223J} used a value of 650 Draine field units  \citep{1978ApJS...36..595D}, or $G_0 = 1100$ in Habing field units \citep{1968BAN....19..421H}.
Here, an edge-on orientation was assumed, meaning that the distance between the tip of IC\,63 and $\gamma$~Cas is taken to be equal to the projected distance of 1.3 pc.
In the work of \citet{2018A&A...619A.170A}, a value of $G_0 \sim 150$ is obtained instead, based on measurements of the FIR emission.
This implies that IC\,63 might be several times further away from $\gamma$~Cas than the projected distance, and that the orientation is not truly edge-on.}
\edit1{The whole cloud has a projected size of $\sim$ 0.5 pc, and the area studied in this work, the tip, is $\sim 0.1$ pc wide.
  IC\,63 is still a dense PDR (\SI{1.2e4}{\per\cubic\cm}), but it has a relatively low column density (\SI{2.3e20}{\per\square\cm}; \citealt{2018A&A...619A.170A}).
Compared to the Horsehead nebula or the Orion bar, this is about an order of magnitude lower in both number and column density, making IC\,63 sufficiently transparent to detect background stars.}

In this work we aim to study the spatial variations of the dust properties observed through extinction, in particular through the $A_V$ and $R_V$ parameters \citep{1989ApJ...345..245C}. IC\,63
is a most suitable target for this, as it has many observable background stars, each of which
provides information about the medium along a specific line of sight. Grouping these stars into
spatial bins makes it possible to measure the average extinction for a number of regions on
the sky. To measure the extinction for each star, we use the same approach as in \citet{2016ApJ...826..104G} for the Panchromatic Hubble Andromeda Treasury (PHAT) survey data \citep{2012ApJS..200...18D}. A catalog of point sources is generated from broadband Hubble Space Telescope (HST)
observations, to which our Bayesian Extinction And Stellar fitting Tool \citep[BEAST, ][]{2016ApJ...826..104G} is applied.

In section \ref{sec:method}, we describe our 7-band photometric observations with HST, and how we extract photometric measurements for 520 background sources. Our Bayesian extinction fitting tool is introduced, and some necessary modifications to it are explained. Section \ref{sec:results} presents the individual fit results for
the sources, and how these were processed to create $A_V$ and $R_V$ maps. In
section \ref{sec:discussion}, we compare our findings to earlier studies of IC\,63, and use data from \textit{Herschel} and \textit{Spitzer} combined with our maps and modified
blackbody fits to derive the $A_V$-normalized dust surface brightness and dust optical depth.
\edit1{To conclude, we present some simple per-pixel fits to the FIR SED, and compare some of the results for IC\,63 with the Horsehead nebula and NGC\,7023.}

\section{Data and Analysis} \label{sec:method}
\subsection{Hubble Observations}
The photometric data for the various sources behind IC\,63 were obtained through observations with the Wide Field Camera 3 (WFC3) of HST, using both the UVIS (UV and
visual) and IR (infrared) channels. We obtained photometric images in the F275W, F336W, F475W, F625W, F814W bands with the UVIS chip, and in the F110W and F160W bands with the IR chip.
This set of \edit1{seven} broadband filters was chosen \edit1{to} cover the stellar SED from the UV to the near infrared, and to optimize the extraction of the dust extinction parameters through SED fitting. The IR measurements provide a way to resolve the degeneracy between the stellar surface temperature and the reddening, while the UV filters constrain the type of extinction. A similar observing strategy, and a more detailed reasoning for choosing each filter can be found in \citet{2012ApJS..200...18D}.

\begin{figure*}[t]
  \centering
  \plotone{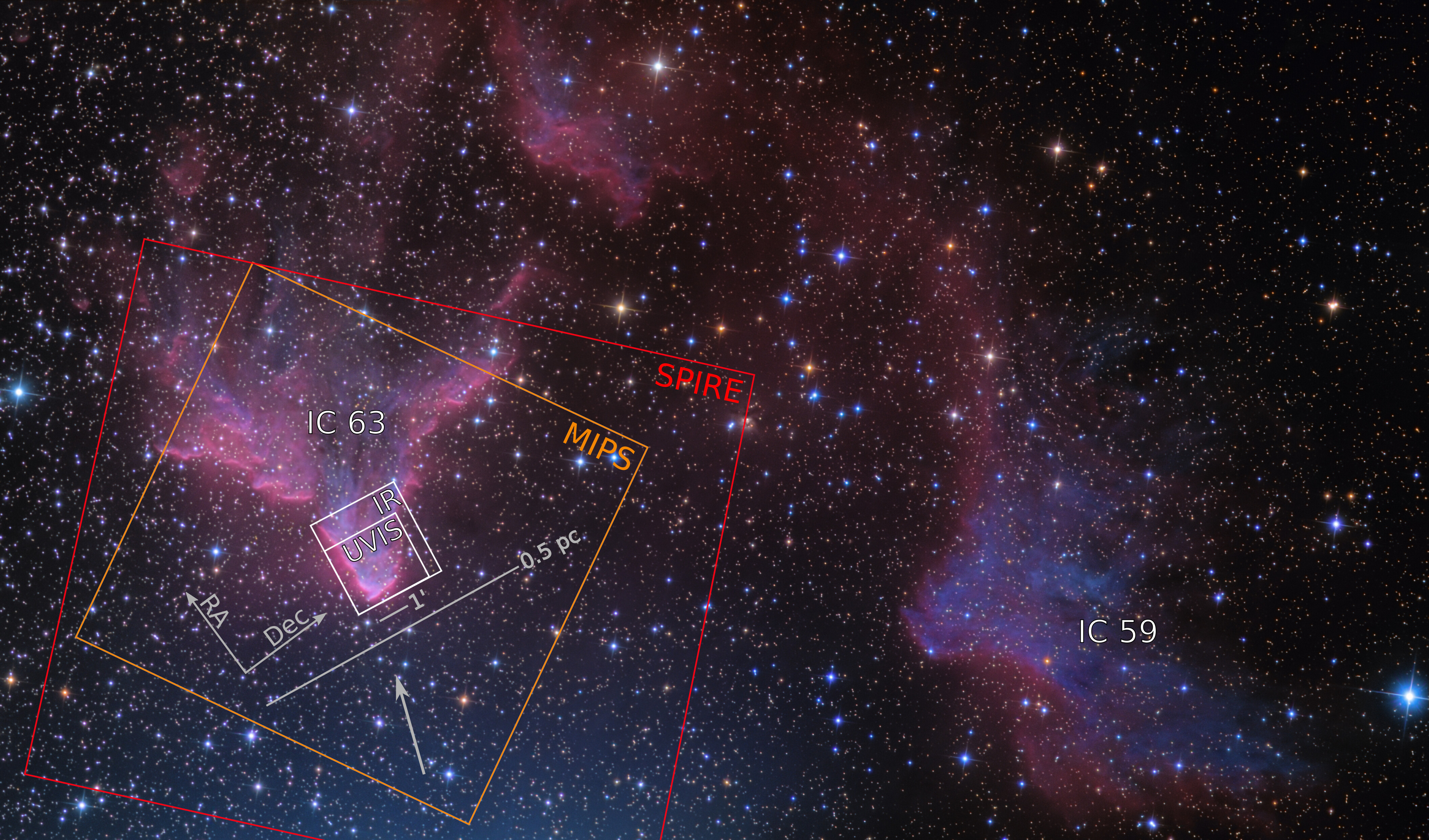}
  \caption{Footprints of the observed data and retrieved archival data.
    The image (credit: Ken Crawford) shows IC\,63 and its environment, including the nearby nebula IC\,59, \edit1{using a combination of exposures in blue, green, red, H$\alpha$, and S\textsc{ii} filters.}
    Between IC\,63 and IC\,59, a structure of dark dusty clouds can be observed, which appears more clearly in FIR images.
    The gray arrow indicates the direction of the radiation from $\gamma$ Cas.
    \edit1{The physical length scale shown assumes a distance of 190 pc.}}
  \label{fig:footprint}
\end{figure*}

The UVIS images span a field of view (FOV) of 162 by 162 arcsec, while the FOV for the IR images is somewhat smaller, at 123 by 136
arcsec (Figure \ref{fig:footprint}).
The observing program consisted of two visits, each two orbits long. During the first visit \edit1{(2016 August 26)}, the F275W and F336W exposures were taken, followed by F814W
and F160W.
During the second visit \edit1{(2016 August 31)}, the F625W, F110W, F475W and additional F275W exposures took place. See Table \ref{tab:proposal} for the exposure times.
The time difference of several days between the visits is intentional, to prevent persistence for the IR exposures.

Guard exposures for bright stars where taken for 5s in F475W and F814W between the two orbits of the first and second visit, \edit2{respectively}. To help dealing with macroscopic features of the detector and cosmic rays, a gap line dither pattern was used \edit1{for} all exposures with UVIS except the guard exposures. For the IR, a small line pattern was used instead to minimize blurring due to IR persistence.
A post-flash was used for all UVIS exposures, except F625W.

\edit1{For all figures that make use of this HST data (such as Figure \ref{fig:collage}), the \texttt{drz} images were used.
These images are created by the standard ``Drizzle'' algorithm of the HST pipeline \cite{2002PASP..114..144F}, which aligns and combines the dithered exposures, removes cosmic rays and artifacts, and corrects for geometric distortion.}
The photometry described in section \ref{sub:photometry} works with the \texttt{flt} files (for IR) and \texttt{flc} files (for UVIS).
\edit1{These contain the individual calibrated exposures.
  The \texttt{flc} files in particular have been corrected for the Charge Transfer Efficiency (CTE) of the detector.
  See the WFC3 data handbook.\footnote{\url{http://www.stsci.edu/hst/instrumentation/wfc3/documentation}}}

\begin{deluxetable}{rrc}
  \tablecaption{Filters and exposure times as listed in the observing program \citep{2015hst..prop14186A}.}
  \label{tab:proposal}
  \tablehead{\colhead{channel and filter} & \colhead{$\lambda_{\text{eff}}$}& \colhead{exposure time}}
  \startdata
    UVIS F275W       & \SI{275}{\nano\meter}   & $2 \times \SI{600}{\s} + 2 \times \SI{349}{\s}$ \\
    UVIS F336W       & \SI{335}{\nano\meter}   & $2 \times \SI{650}{\s}$             \\
    UVIS F475W       & \SI{475}{\nano\meter}   & $4 \times \SI{450}{\s}$        \\
    UVIS F625W       & \SI{625}{\nano\meter}   & $4 \times \SI{420}{\s}$            \\
    UVIS F814W       & \SI{814}{\nano\meter}   & $4 \times \SI{449}{\s}$            \\
    IR F110W         & \SI{1.10}{\micro\meter} & $2 \times \SI{350}{\s}$            \\
    IR F160W         & \SI{1.60}{\micro\meter} & $2 \times \SI{400}{\s}$
    \enddata
\end{deluxetable}

\subsection{Ancillary Data}
\begin{figure*}[tb]
  \centering
  \plotone{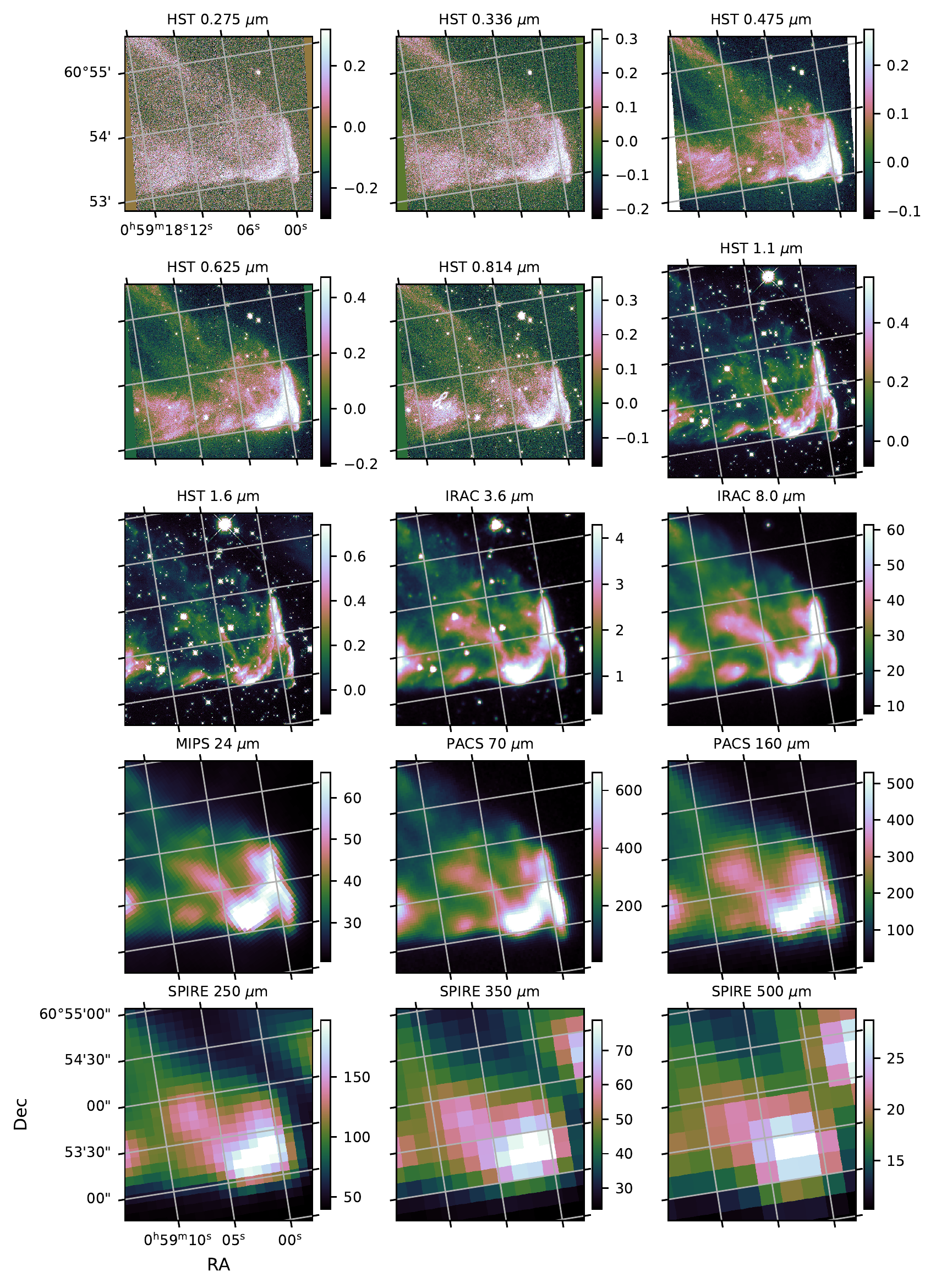}
  \caption{Overview of the stacked and drizzled HST images, and a selection of archived imaging
    data at longer wavelengths. The IRAC 4.5 and 5.8 micron images are not shown.
    The archived data has been reprojected onto the coordinate frame
    of the HST UVIS images. \edit1{The color scales show the flux in MJy sr$^{-1}$.}}
  \label{fig:collage}
\end{figure*}

The FIR PACS \citep{2010A&A...518L...2P} and SPIRE \citep{2010A&A...518L...3G} imaging data were obtained from the \textit{Herschel} Science Archive. They were originally part of
a key program for the \textit{Herschel} Space Observatory \citep{2010A&A...518L...1P},
``Evolution of Interstellar Dust'' \citep{2010A&A...518L..96A}.
We also retrieved IRAC \citep{2004ApJS..154...10F} and MIPS \citep{2004ApJS..154...25R}
photometric data from the \textit{Spitzer} heritage archive,
taken under the ``Star Formation in Bright Rimmed Clouds'' program (ID 202).
An overview of the images can be found in Figure \ref{fig:collage}, where these data were
reprojected onto the same coordinate frame as the drizzled HST UVIS
images.

\subsection{Point Source Photometry} \label{sub:photometry}
\subsubsection{Source Extraction}
We employ a technique analogous to the one used to extract similar photometric catalogs
from the PHAT \citep{2012ApJS..200...18D, 2014ApJS..215....9W}, SMIDGE \citep{2017ApJ...847..102Y} and METAL \citep{2019arXiv190106027R} data. Following a precise astrometry and alignment step, the photometry software
DOLPHOT \citep{2016ascl.soft08013D} is used to automatically detect the point sources, and to
determine their positions and fluxes, based on Point Spread Function (PSF) fitting.
The details about this photometric extraction routine are described in \citet{2014ApJS..215....9W}.
However, we do use the same tweaks as \citet{2019arXiv190106027R}. The \edit1{Tiny Tim PSF libraries \citep{2011SPIE.8127E..0JK}} are used instead of the Anderson libraries \edit1{\citep{2000PASP..112.1360A}}. \edit1{Additionally, for UVIS, we make use of the \texttt{flc} images, which} have already been corrected for CTE.
Therefore, a separate CTE correction step after the PSF fitting is no longer necessary.

\subsubsection{Removal of Spurious Detections}
\label{subsub:culling}
\begin{figure}[t]
  \plotone{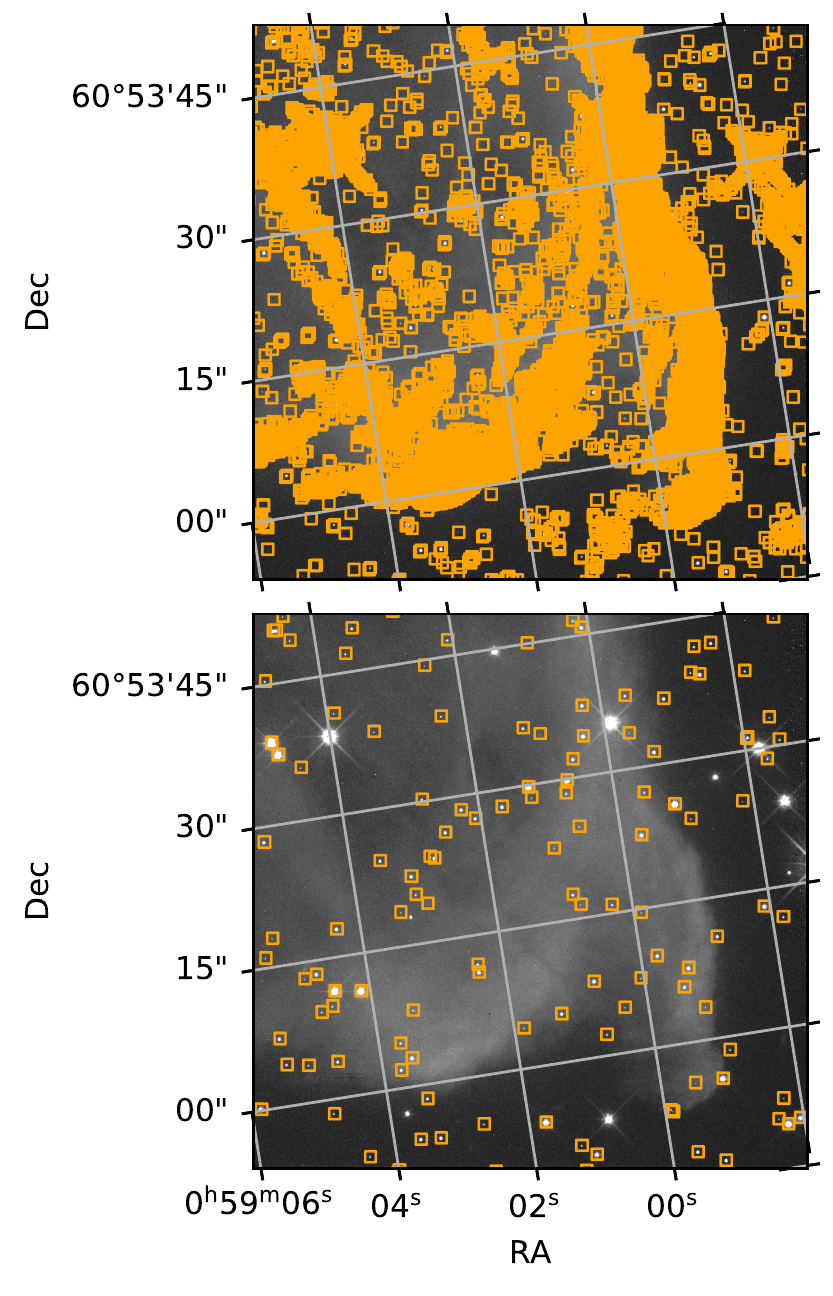}
  \caption{\edit1{Illustration of the cleaning step, using a zoom-in on the tip of the nebula.} The orange boxes indicate the positions of detected sources in our catalog, before \edit1{(\textit{top})} and after \edit1{(\textit{bottom})} applying the criteria described in section \ref{subsub:culling}. All of the detections due to the extended emission are removed, while only a couple of the stars are missed.}
  \label{fig:spurious}
\end{figure}

Due to the relatively bright extended emission in IC\,63 (stellar light scattered by dust, gas emission lines due to recombination), the majority of the sources listed by the automatic
source detection algorithm mentioned above are false positives. This can be seen in
Figure \ref{fig:spurious}, where most of the detections clearly coincide with the extended
emission. Fortunately, we found some simple
criteria to separate the bulk of these spurious detections from the real point sources, making
use of some of the quantities provided by the PSF fitting routine. These quantities are given in
certain columns of the photometric catalog.

The first criterion is a cut on the relative flux error in the F814W band. We only keep the sources
for which the error to flux ratio $\texttt{F814W\_ERR} < 0.08$ \edit1{($\text{SNR} > 12.5$)}. This excludes almost all of the entries due to the
extended emission and some diffraction spikes of bright stars.

Secondly, an extra cut on the crowding is performed, specifically
$\texttt{F814W\_CROWD} < 0.25$. This removes a handful of overlapping sources, as well as
several remaining detections located at diffraction spikes.
The thresholds for these first two cuts were obtained through some trial and error.
By visually inspecting plots such as Figure \ref{fig:spurious}, we found that these values remove most of the spurious detections, while keeping detections of obvious point sources intact.

Lastly, we remove all sources that are outside the region where observations exist in all bands, to provide a homogeneous dataset for analysis. The right panels of Figure \ref{fig:spurious} show the positions of the sources that are left after applying the cuts.
There are 520 sources remaining, over the area covered by both the UVIS and IR chip.

\subsection{Physics Model}
To model the stellar SED attenuated by dust for each source, we
employ the Bayesian Extinction And Stellar Tool \citep[BEAST;][]{2016ApJ...826..104G}. This tool works with a 7D grid of SED
models, one dimension for each parameter ($M_\text{ini}$, $t$, $Z$, $d$, $A_V$, $R_V$, $f_A$; see description below and Table \ref{tab:grid}). For each set of parameters, the
physics model makes a prediction for the observed SED of the source. This is done by first
constructing models for the stellar spectrum and the extinction curve. The model spectrum is
then extinguished according to this curve at each of its wavelength points, and integrated over
the transmission curve of each filter.

\subsubsection{Stellar Parametrization}
In the current implementation, the models for the stellar
spectra are calculated based on \edit1{four} parameters, specifying the star's birth mass \edit1{$M_\text{ini}$}, age $t$, metallicity $Z$, and distance $d$.
Starting from the first three parameters, the luminosity of the star and the shape of the
spectrum are calculated using a combination of publicly available stellar atmosphere grids \citep{2003IAUS..210P.A20C, 2003ApJS..146..417L, 2007ApJS..169...83L} and
evolutionary tracks \citep{2008A&A...482..883M, 2010ApJ...724.1030G, 2012MNRAS.427..127B, 2013EPJWC..4303001B}. The reasoning behind this choice of parameters can be
found in \citet{2016ApJ...826..104G}.
\edit1{The predicted fluxes generated by this model are purely synthetic}.

\subsubsection{New BEAST Feature: Distance as an Extra Stellar Parameter}
Prior to this work only a single value for the distance was supported, as this tool had only been applied in the extragalactic context where a single distance to all stellar sources can be assumed. For IC\,63, setting
a single distance is of course not possible, as all the background stars are located within the
Galaxy, and hence they have a wide variety of distances. Therefore, in the version of the BEAST used for this work, the distance was implemented as an extra parameter for the stellar model.

\subsubsection{Extinction Parametrization}
The other \edit1{three} parameters describe the shape and magnitude of the extinction curve. The BEAST
features a two-component dust extinction model, mixing two different shapes (wavelength dependencies);
the type-$\mathcal{A}$ curve, which models the average extinction
as measured in the Milky Way (MW), and the type-$\mathcal{B}$ curve, which models that of the
Small Magellanic Cloud (SMC) Bar \citep{2003ApJ...594..279G}.
For the MW-like type, the shape of the extinction curve
$A_\lambda / A_V$ depends on $R_V$
\begin{equation}
 \frac{A_\lambda}{A_V} = k_{\lambda, \mathcal{A}}(R_V)
\end{equation}
where $A_V$ is the total Johnson V-band extinction by the dust, and $R_V = A_V / E(B - V)$. This $R_V$ dependence is modeled according to \citet{1999PASP..111...63F}. The SMC-like extinction model $k_{\lambda, \mathcal{B}}$ does not depend on $R_V$, and its wavelength dependence is modeled using an average of the measurements given by \cite{2003ApJ...594..279G}.
Within this model, the value of $A_V$ is a way to
express the total dust column, while \edit1{an increase in $R_V$ is believed to reflect a shift of the grain size distribution towards larger sizes \citep{1989ApJ...345..245C}.}
A third
parameter, $f_\mathcal{A}$, takes a linear combination of shapes $\mathcal{A}$ and
$\mathcal{B}$:
\begin{equation}
  k_\lambda (R_V, f_{\mathcal{A}}) = f_{\mathcal{A}} k_{\lambda, \mathcal{A}}(R_V) + (1 - f_\mathcal{A}) k_{\lambda, \mathcal{B}}
\end{equation}

\edit1{Note that this extra degree of freedom is necessary because the type-$\mathcal{A}$ and type-$\mathcal{B}$ curves only describe the \textit{average} MW and SMC extinction.
  For individual sightlines, both components are needed to model the scatter on the $R_V$-dependent relationship of \citet{1989ApJ...345..245C}.
There are certain sightlines in the MW that exhibit SMC-like extinction in the UV \citep{2003ApJ...598..369V}, as well as sightlines in the SMC with MW-like extinction \citep{2003ApJ...594..279G}.}

This extinction model is then applied to the stellar spectra generated from the first four parameters.

\subsubsection{Model Details for This Work}
The details about the parameter ranges and their spacing in the BEAST physics model grid are summarized in Table
\ref{tab:grid}. Suitable ranges and resolutions were obtained through a combination of previous
experience (such as Table 1 in \citealt{2016ApJ...826..104G}), some trial and error, and some
compromises considering computational resources.
\edit1{Assuming a distance to IC\,63 of about 190 pc, we initially considered a lower bound of 150 pc for the new distance parameter.
A test run for the fitting showed that only a handful of sources were closer than 500 pc.
Therefore, we decided to change the lower bound for the distance to 500 pc, providing a slight increase in resolution.
Analogously, we confirmed that an upper bound of 15000 pc was sufficient to fit all observed stars.}

The parameters shown in the table give rise to about \SI{5.3e9}{} models, and without
compression, about 1.6 TB of disk space is needed to store this grid.
The resolution for most of the parameters is
relatively low, because a relatively large range and number of bins is needed for the distance.
\edit1{For our purposes however, a relatively rough estimate of $A_V$ and $R_V$ suffices, since we only need the average values over relatively large pixels.
  Within such a pixel, we found the range of $A_V$ and $R_V$ values to be broader than the chosen resolution.
  This spread can either be caused by the different stellar distances resulting in a different contribution by the diffuse Galactic ISM, or by small scale variations in the cloud.
  The spread on $A_V$ and $R_V$ within each pixel will be used to estimate the error the mean values.
In any case, the precision on the $A_V$ and $R_V$ maps is mostly limited by the number of sources and the chosen pixel size, and not by the precision of individual $A_V$ or $R_V$ measurements.}

\begin{deluxetable*}{lclllc}
  \tablecaption{Specification of the parameter grid for the physics model. \label{tab:grid}}
  \tablehead{\colhead{Parameter} & \colhead{Description} & \colhead{Min} & \colhead{Max} &
    \colhead{Resolution} & \colhead{Prior}} \startdata
  $\log(t)$ (years)              & stellar age             & 6.0           & 10.13         & 0.3                       & constant SFR \\
  $\log(M_\text{ini})$ ($M_{\astrosun}$)    & stellar mass            & -1.0          & 2.5           & variable\tablenotemark{a} & Kroupa IMF   \\
  $\log(Z)$ (mass fraction)      & stellar metallicity     & -3            & -1.2          & 0.15                      & flat in $Z$ \\
  $\log(d)$ (pc)                 & distance                & 2.7           & 4.2           & 0.037\tablenotemark{b}    & flat in $\log(d)$        \\
  $A_V$ (mag)                    & dust column             & 0             & 10.05         & 0.08                      & flat         \\
  $R_V$                          & dust average grain size & 1.5           & 5.5           & 0.5                       & flat         \\
  $f_\mathcal{A}$ & dust mixing parameter & 0 & 1 & 0.2 & flat \enddata
  \tablenotetext{a}{Supplied by isochrone model grid}
  \tablenotetext{b}{This corresponds to 40 logarithmic steps from 500 to 15000 pc}
\end{deluxetable*}

The priors are shown in the rightmost column of Table \ref{tab:grid}. For the new distance parameter, the prior is chosen to be flat as a function of $\log(d)$.
As a consequence, for some stars a degeneracy exists between the estimated distance and mass (or
luminosity): an observed star can either be far away and of a luminous type, or
nearby and of a fainter type. The multi-band photometry partly resolves this degeneracy, and we did not find this effect to be problematic for constraining the extinction.

\begin{figure}
  \centering
  \plotone{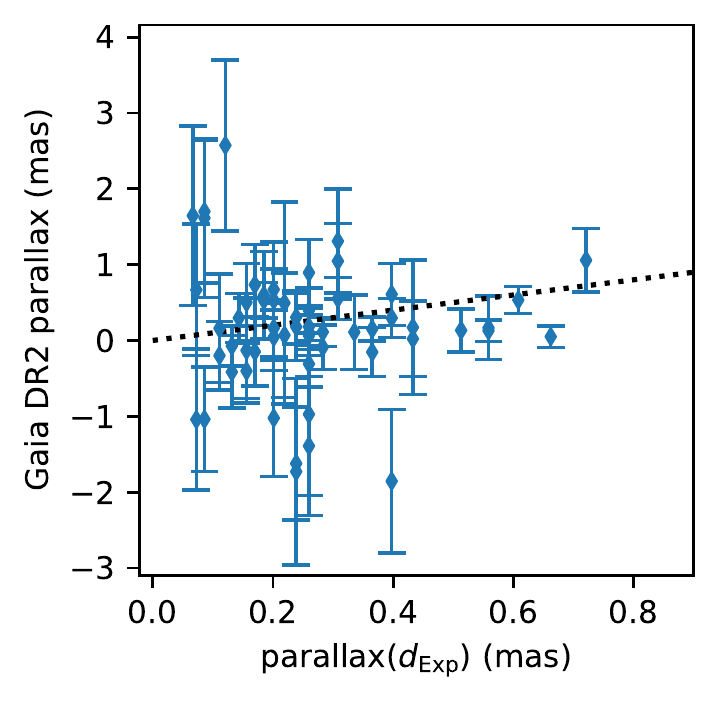}
  \caption{\edit1{Comparison of the Gaia DR2 parallaxes and the parallax derived from the distance expectation values $d_\text{exp}$, for the 59 sources present in both catalogs.}}
  \label{fig:gaia}
\end{figure}

In the future, more physically motivated priors could be implemented for the distance, such as an exponentially decreasing space density \citep{Bailer-Jones2018}.
Another option would be to make direct use of the parallaxes provided by Gaia DR2 \citep{ 2016A&A...595A...1G, Gaia-Collaboration2018, Luri2018}, and use them as source-specific priors in a post processing step.
However, since there are only 59 out of $\sim$500 stars in our data set with Gaia data available, this would have a limited impact on the results of this work.
Moreover, we did cross check the Gaia data with our fit results, and found that the parallaxes derived from the distances were consistent with the fairly uncertain Gaia parallaxes (Figure \ref{fig:gaia})

\subsection{Noise Model} \label{sub:noise}
\subsubsection{Artificial Star Tests}
By applying the physics model to all combinations of the \edit1{seven} parameters, a grid of theoretical
SEDs is constructed, representing the observations in the relevant filters under perfect
circumstances. But in a realistic setting, when the flux is measured from imaging data, the
flux values that are extracted for a certain point source are not only affected by shot (Poisson) noise and the PSF, but also by nearby objects, various sources of background, instrumental artifacts, and the extraction algorithm itself.
Nearby point sources can create \textit{crowding} effects \citep{2016ApJ...826..104G}, which increase the noise and cause
an overestimation of the flux. However, since the density of observed stars in IC\,63 is very low compared to
the size of the PSF, these effects are negligible. For IC\,63, the main expected contributor to the
observation model is the presence of a foreground and background of extended emission, such as H$\alpha$
emission of the gas or stellar light \edit1{scattered} by dust grains.

To quantify how significantly one of the model SEDs differs from the observed SED of a certain star, we create a model describing the uncertainty (error $\sigma$) and systematic deviation (bias $\mu$) for each flux. While the photometric catalog produced by the PSF fitting does contain values for
the uncertainty on each flux (which we used to perform the input cleaning step), these do not
cover all possible error sources. Instead, we treat the errors as a property of each theoretical
model: for each grid point $i$, we aim to calculate the error or uncertainty $\sigma_i$ and a
bias $\mu_i$ in every band.

The starting point for this calculation is a set of Artificial Star Tests (ASTs). The input for
a single AST consists of a theoretical SED and a position. These parameters are used as input
for the fake star mode of DOLPHOT. This routine will insert a fake star with the given SED into
each of our observations, simulate the PSF and photon noise, and then perform the exact same photometric extraction routine. The result
is a set of output fluxes, representing a mock-observation of a point source with that specific
theoretical SED. By putting the same source at many different positions, statistical information
is obtained about the deviations that occur when observing that specific SED $i$, from which
appropriate values for $\sigma_i$ and $\mu_i$ can be derived.

Ideally, such a set of tests would be performed for each SED $i$ in the physics grid, leading to
an individual measurement of the noise for each model. However, performing a single AST takes
about 2 minutes of computing time on a modern processor, so doing this for the billions of models contained in the physics grid is practically impossible. Therefore, it is
customary to produce a sample of SEDs that is representative of all the models relevant for
fitting the observed sources.

\subsubsection{New BEAST Feature: Uniform SED Sampling by Magnitude}
Since the distance parameter can
strongly scale the model SEDs, faint stars at long distances and luminous stars at short
distances will have model fluxes that are far away from the minimum and maximum of the observed
catalog. Their noise parameters are not useful for fitting our observed SEDs. Therefore we only
consider models that have fluxes that fall between the minimum and maximum of the observed SEDs, with 1 magnitude of
leeway to allow for a range of uncertainty. The magnitude ranges of the remaining models are
then discretized into 25 bins for each filter, and SEDs are chosen randomly from the physics
grid and added to the AST input list. For each SED that is chosen, the corresponding magnitude
is determined in each band, so we can keep track of the number of ASTs that will cover each
filter-mag bin. The algorithm continues until we have 50 samples in each bin, and the resulting set
of AST input SEDs will have a magnitude distribution that is more or less flat in each filter.

\begin{figure}
    \centering
    \plotone{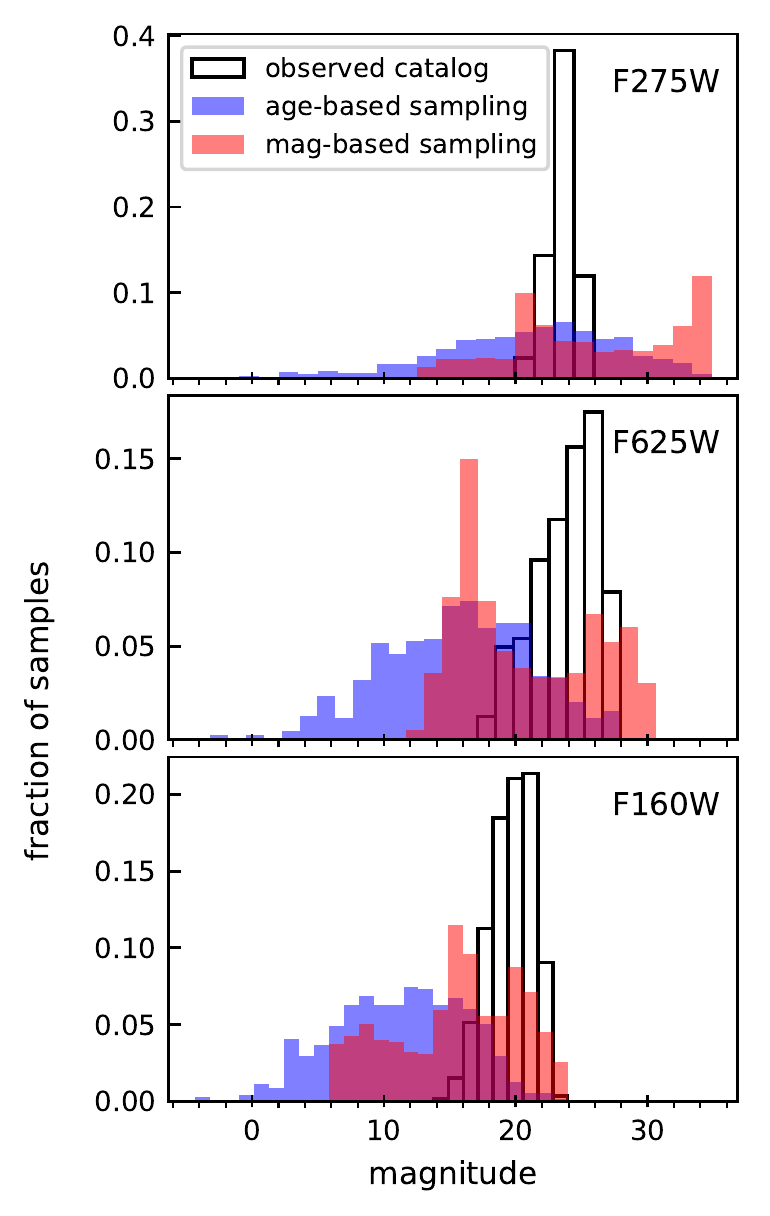}
    \caption{\textit{Red:} Magnitude distribution produced by the new sampling method, which aims to provide a minimum number of samples for each magnitude interval. \textit{Blue:} The old sampling method, which simply picks a fixed number of SED models from each stellar age bin. \textit{Black outline:} Magnitude distribution of the observed catalog (only fluxes > 0 shown).}
    \label{fig:sedsampling}
\end{figure}

Figure \ref{fig:sedsampling} shows the difference between the original method which evenly samples by stellar age, and the new method which evenly samples by magnitude. To show the magnitudes that need to be covered, the distribution of the observed catalog is also shown. The original method does not sufficiently sample the low brightness range, while simultaneously producing many samples that are several magnitudes brighter than the maximum of the observed catalog. As designed, the new method produces a certain minimum number of samples for each flux bin, within a more suitable range. The peaks that appear in the distribution depend on the contents of the physics grid and the brightness cutoffs.
\edit1{The extra stars that populate those peaks} were picked to fill the rarer flux ranges in other filters. Note that the new method still produces many samples outside of the observed flux range, because we only require that at least 3 bands fall within that range. This allows for a larger variety of SED shapes to be picked for the ASTs.

Using this approach, a list of about 2500 SEDs was generated. This list is duplicated multiple times while giving a random position to each SED, leading to 180000 unique ASTs.
The list of ASTs was split up into a set of jobs, each of which have a manageable runtime. These jobs were executed across three
machines with the help of GNU parallel \citep{Tange2011a}, using 78 of the available cores over the course of 3 days. The results were then merged and processed into a single \textit{fake star catalog}, of the
same format as the one containing the observed photometry.

\subsubsection{From ASTs to Noise Model} \label{subsub:noisemodel}
Before the fake star catalog is further processed by the BEAST to generate the noise model, the
same selection function used in section \ref{subsub:culling} is applied. This way it is ensured that the observation model generates a set of noise parameters that is representative of our cleaned ensemble of observed stars.
Note that this also removes ASTs which are missing a flux in one or more bands, which includes
those positioned outside the IR chip's FOV.
After this step, about 80000 fake stars remain.
\edit1{Although a large fraction (100000 out of 180000) of the fake stars ended up either undetected or rejected, we confirmed that the remaining catalog still sufficiently covers the required magnitude range, as in Figure \ref{fig:sedsampling}.}

\begin{figure}[t]
  \centering
  \plotone{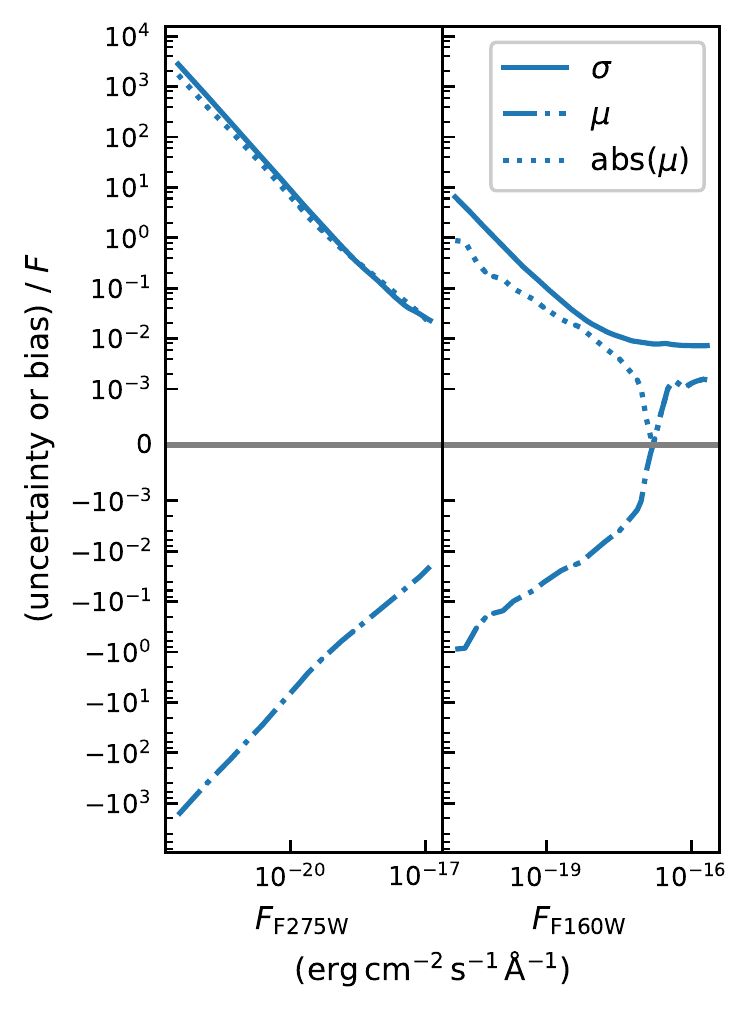}
  \caption{Flux-dependent model of the bias $\mu$ and \edit1{uncertainty} $\sigma$ \edit1{for the shortest and longest wavelengths}.
    The vertical axis is a symmetrical log scale (linear between \edit1{\num{-e-3}} and \edit1{\num{e-3}}), which shows the sign change of $\mu$.
    The absolute value of $\mu$ is also plotted, showing that the bias correction is within $1\sigma$ almost everywhere.}
  \label{fig:toothpick}
\end{figure}

To generate the noise model for each SED in the physics grid, the BEAST's most conservative method is used, which considers each flux individually and as such ignores any correlations between the bands.
This method minimizes the number of ASTs needed, which is already quite large due \edit1{to} the broad range of fluxes needed to model all the distances.
Similarly to how the
AST input SEDs were chosen, the flux ranges of the model SEDs are divided into 30 \edit1{logarithmic} bins. For each
bin, the input and output SEDs of the relevant ASTs are compared, producing statistical values
for the error and the bias in that flux range, for a specific filter. By interpolating between
these bin-averaged values, each filter now has a model that predicts $\sigma$ and $\mu$ as a
function of the theoretical flux $F$. This behavior is shown in Figure \ref{fig:toothpick}.

One of the main features of the noise model is that the bias seems to be negative for all but the brightest fluxes, in all bands. This means that the fluxes extracted by the photometry routine are on average smaller than the input fluxes of the fake stars. \edit1{To take this effect into account, we forward model the observed fluxes by adding the value of $\mu$ to the theoretical model $F$, as shown in equation \ref{eq:delta} of the next section.
In this case specifically, the negative value for $\mu$ makes it possible to fit our model to the many stars with negative flux measurements in the UV. The theoretical model cannot produce negative flux values, but adding the bias shifts these values down to the correct level.}

For weak fluxes, the constant slope of the relative $\sigma$ and $\mu$ implies that the noise parameters are dominated by photon noise and/or constant features in the image. For stronger fluxes, the relative error levels off while the bias becomes positive (but much smaller than the error). This means that fractional deviations dominate instead, which are most likely caused by the PSF or imperfections in the photometry routine. For all bands except the UV band, the absolute value of $\mu$ seems to be smaller than $\sigma$. This means that the exact value of $\mu$ will have less of an effect on the results, as it only causes deviations within $1\sigma$.

Since we suspect that the background/foreground extended emission provides the main contribution to the noise model, we experimented with a noise model that is stratified according to the intensity of this emission. In the appendix, we explain how we determine 4 regions of similar background/foreground intensity, and create a separate noise model for each of them. Ultimately, the fitting results obtained using these individual noise models were not found to be significantly different, so the rest of the analysis was performed using the single noise model presented in this section. However, these tools integrate well with and improve upon existing code that deals with varying levels of crowding, and have therefore been incorporated in the main BEAST branch.

\subsection{Fitting Procedure and Output}
Once the necessary models have been set up, the fitting step comes down to calculating a 7D
posterior probability distribution
\begin{equation}
P(\theta | F_\text{obs}) \propto P(F_\text{obs} | F(\theta), \Sigma(\theta)) P(\theta)
\end{equation}
for each of the observed point sources. Here, $F_\text{obs}$ is the observed SED, and
$F(\theta)$ and $\Sigma(\theta)$ represent the SED and the noise parameters for one of the
models, evaluated at a specific point $\theta$ of the 7D parameter grid.
\edit1{The different priors listed in Table \ref{tab:grid} are multiplied to form the 7D prior $P(\theta)$}.
The grid point
$\theta$ at the global maximum of \edit1{the posterior} distribution is considered to be the best fitting model.
The likelihood function is a 7D multivariate Gaussian:
\begin{gather}
  P(F_\text{obs} | F(\theta), \Sigma(\theta)) = \frac{1}{\mathcal{N}} e^{-\chi^2 / 2}                \\
  \label{eq:chi2} \text{with} \quad \chi^2 = \Delta(\theta)^T \mathbb{C}(\theta)^{-1} \Delta(\theta) \\
 \label{eq:delta} \text{and} \quad \Delta(\theta) = F_\text{obs} - \mu(\theta) - F(\theta)
\end{gather}
\edit1{where the dimension $\mathcal{N} = 7$ equals the number of fluxes, and $\mathbb{C}(\theta) = \text{diag}(\sigma(\theta))$ and $\mu(\theta)$ are the covariance matrix and the bias vector.
  The values for $\sigma(\theta)$ and $\mu(\theta)$ are calculated as a function of $F(\theta)$, as described in section \ref{subsub:noisemodel}.
  Note that $\mathbb{C}$ is diagonal for this work, since this calculation is done for each band individually.}

In principle, the whole 7D log-likelihood can be written to disk for each star. In practice,
only a local area around the peak of the posterior distribution is written out. Additionally,
the 1D marginal probability distributions for each parameter are also available. For our
purposes however, some statistics derived from the posterior are sufficient. For all the parameters, the
best fit, the expectation values, and the 16th, 50th and 84 percentiles are calculated and
stored in a table that contains one row per source. The $\chi^2$ of the least-squares model and
$\ln P$ of the best fitting model are also stored. The former gives us some insight about the
quality of the grid and the noise model, and the source itself. A high $\chi^2$ usually means that the grid
range or resolution is not sufficient, that there might be something wrong with the noise model, or that a source in particular has an SED that is very
much unlike that of an extinguished star.

\section{Results} \label{sec:results}
\subsection{Postprocessing} \label{sub:postprocess}
\begin{figure*}[htbp]
  \centering
  \plotone{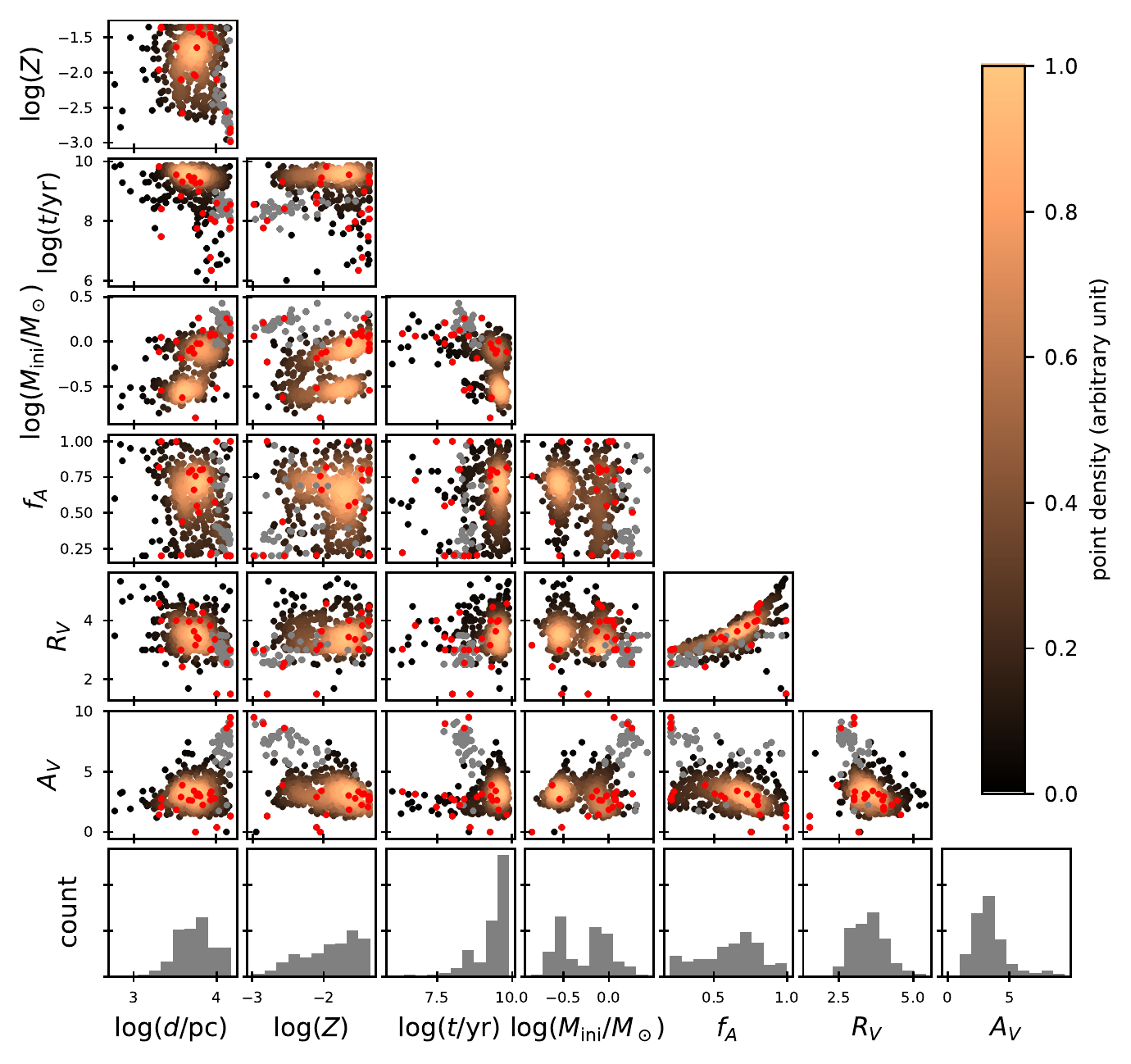}
  \caption{Distribution and correlation of the expectation values for the fit parameters, of all
    520 sources.
    \edit1{The density of the points is represented by the color scale.
      \textit{Red:} stars with $\chi^2 > 100$.
      \textit{Gray:} stars with $\log T > 3.78$.
      After applying these cuts, 462 stars remain.}}
  \label{fig:corner}
\end{figure*}

\begin{figure}[htbp]
  \centering
  \plotone{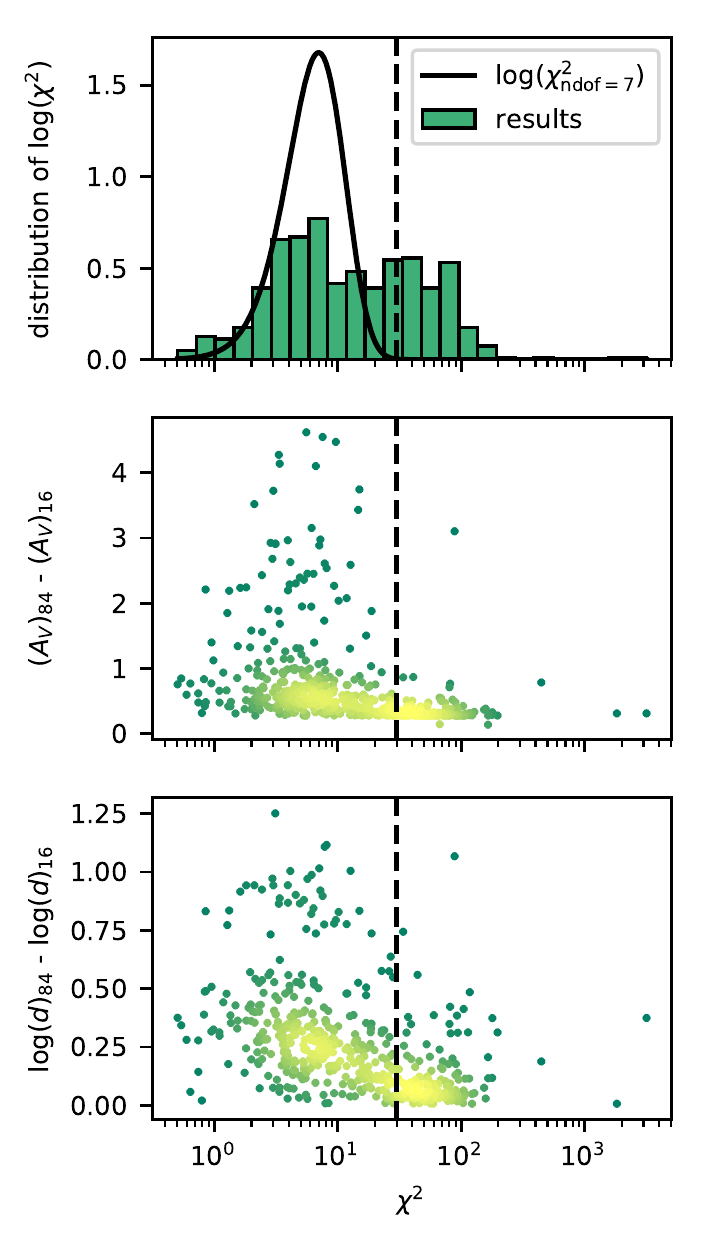}
  \caption{\textit{Top panel:} $\chi^2$ histogram for the fit results and the theoretical distribution for $\log(\chi^2)$ that would naively be expected for 7 degrees of freedom.
  \textit{Bottom two panels:} Relationship between $\chi^2$ and the width of the 1D posterior distributions for $A_V$ and $\log(d)$ (defined by their 84th and 16th percentiles). The color scale visualizes the the density of overlapping points on the plot. The dashed line indicates $\chi^2 = 30$. Almost all sources that have a $\chi^2$ higher than 30 have narrow posterior distributions for $A_V$, $\log(d)$, or both.}
  \label{fig:chi2}
\end{figure}

Following the workflow described in the previous section, we obtained a catalog for all
520 sources (Figure \ref{fig:spurious}) that were left after applying the criteria from section
\ref{subsub:culling}. This catalog contains statistics for the seven main parameters, as well as a set of derived stellar quantities:
current mass $M_\text{act}$ \edit1{(derived from age and initial mass $M_\text{ini}$)}, radius $R$, luminosity $\log(L)$, surface gravity $\log(g)$, temperature $\log(T)$ and \edit1{absolute bolometric} magnitude $m_\text{bol}$.
The
distribution of the expectation values of the seven parameters and the correlations between them are
shown in Figure \ref{fig:corner}, and the $\chi^2$ distribution is shown in Figure
\ref{fig:chi2}.
The latter seems to be slightly peaked around 7, which could naively be attributed to the fact that we have measurements in \edit1{seven}
bands. But in practice, the distribution of this $\chi^2$ is not as simple to interpret, because there are significant correlations between the observed fluxes in different bands. The value can only be used as a rough check that the fits are reasonable.

The two bottom panels of \edit1{Figure} \ref{fig:chi2} show that for almost all sources with a higher $\chi^2$ (30 to 200), the width of their 1D posterior $A_V$ and distance distribution is near the minimum cutoff.
This means that the resulting fits are acceptable despite their high $\chi^2$ values; their accuracy is simply limited by the resolution of the grid. Decreasing the $A_V$ and distance spacing will likely improve the $\chi^2$ and make the $A_V$ distributions narrower, but we consider the precision achieved with the current parameters sufficient for the analysis that follows. Since the distance is a nuisance parameter which is integrated over to obtain the $A_V$ and $R_V$ expectation values, increasing the distance resolution will arguably have a negligible effect.

\begin{figure}[htbp]
    \centering
    \plotone{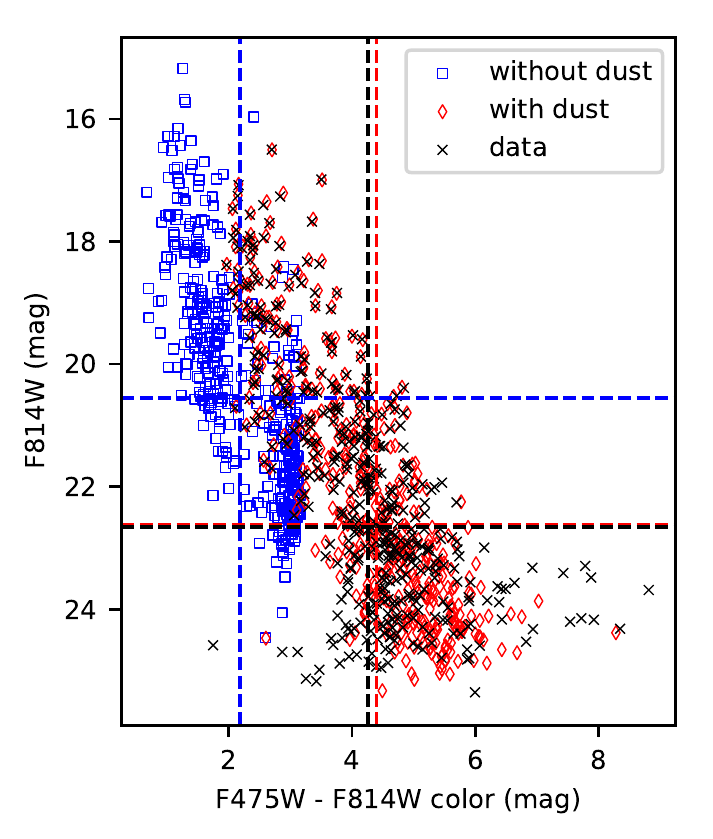}
    \caption{Color magnitude diagram of the (non-negative) raw data \edit1{and the \edit2{expectation} values of the theoretical model fluxes.}
      The dashed lines indicate the median magnitude and color of each set.
      Bright sources seem to have colors and magnitudes that are closely matched, as can be seen from the nearly overlapping crosses and diamonds.
      For weak sources, this overlap disappears due to noise.}
    \label{fig:cmd}
\end{figure}

Based on the corner plot of Figure \ref{fig:corner}, there seem to be no correlations or degeneracies \edit1{which could cause problems.
The $R_V - f_A$ relationship is to be expected, as the type-$\mathcal{B}$ dust extinction component has a fixed $R_V = 2.74$ \citep{2016ApJ...826..104G}; the results with low $f_A$ have an $R_V$ which is closer to this value.}
It is however remarkable that the population of stars seems to have a bimodal
distribution of the fitted initial mass $M_\text{ini}$.
Figure \ref{fig:cmd} shows a color magnitude diagram, and it can be seen that the \edit1{models without \edit2{extinction}} seem to have a gap along the color axis.
\edit1{Upon closer inspection of the derived parameters, the high-mass group centered at around $0.8\,M_{\astrosun}$ has temperatures ranging from 4300 to 6200 K, which corresponds to main sequence stars.
For the low mass group at $0.3\,M_{\astrosun}$, the temperatures range from 2900 to 3600 K, which means it entirely consists of red dwarfs.
Looking at the fitted distances in Figure \ref{fig:corner}, the red dwarfs are only observed from 1000 up to 7300 pc, while the more massive, hotter stars are distributed across the entire range from 1500 to 15000.}

\edit1{To check whether this feature could be produced by two different populations, we created a simulated catalog of stars around the line of sight towards IC\,63, using the Besançon Galaxy model \citep{2003A&A...409..523R}.
When no magnitude limit is used (\texttt{AbsMag\_max} = 99, all observed band limits at 99), the resulting log-mass distribution forms a broad peak between roughly 0.05 and $1.4\,M_{\astrosun}$. 
At the center of this peak ($0.4\,M_{\astrosun}$), a small dip does appear, of about 10\% of the height of the distribution. 
We experimented with some magnitude cuts for the generated catalog, and these generally caused the dip in the log-mass distribution to become less pronounced.
Concluding, it is not out of the question that the observed mass bimodality could be caused by the intrinsic mass distribution along the line of sight. 
While we suspect the observational selection function and possible fitting artifacts to have a significant effect, it remains unclear why exactly there is such a prominent gap in the distribution of the fitted masses.}

We perform a second cleaning step, which removes sources from the output catalog for which the
obtained parameters are likely to be incorrect. For certain sources, the shape of the observed
SED does simply not resemble that of a star with a certain extinction. The reason for this can
be physical, because it's a type of star not covered by the BEAST model, or because there is some other phenomenon at play which is not common enough across the image to be captured by the noise model (e.g.
unusually strong scattered light, background galaxy, diffraction spikes). Most of these are easy
to spot, with $\chi^2$ usually being several hundreds or thousands. Based on the histogram shown
in Figure \ref{fig:chi2}, we apply a cutoff of $\chi^2 < 100$ to remove the worst offenders.
\edit1{They have been highlighted in red in Figure \ref{fig:corner}.}

\begin{figure}[htbp]
    \centering
    \plotone{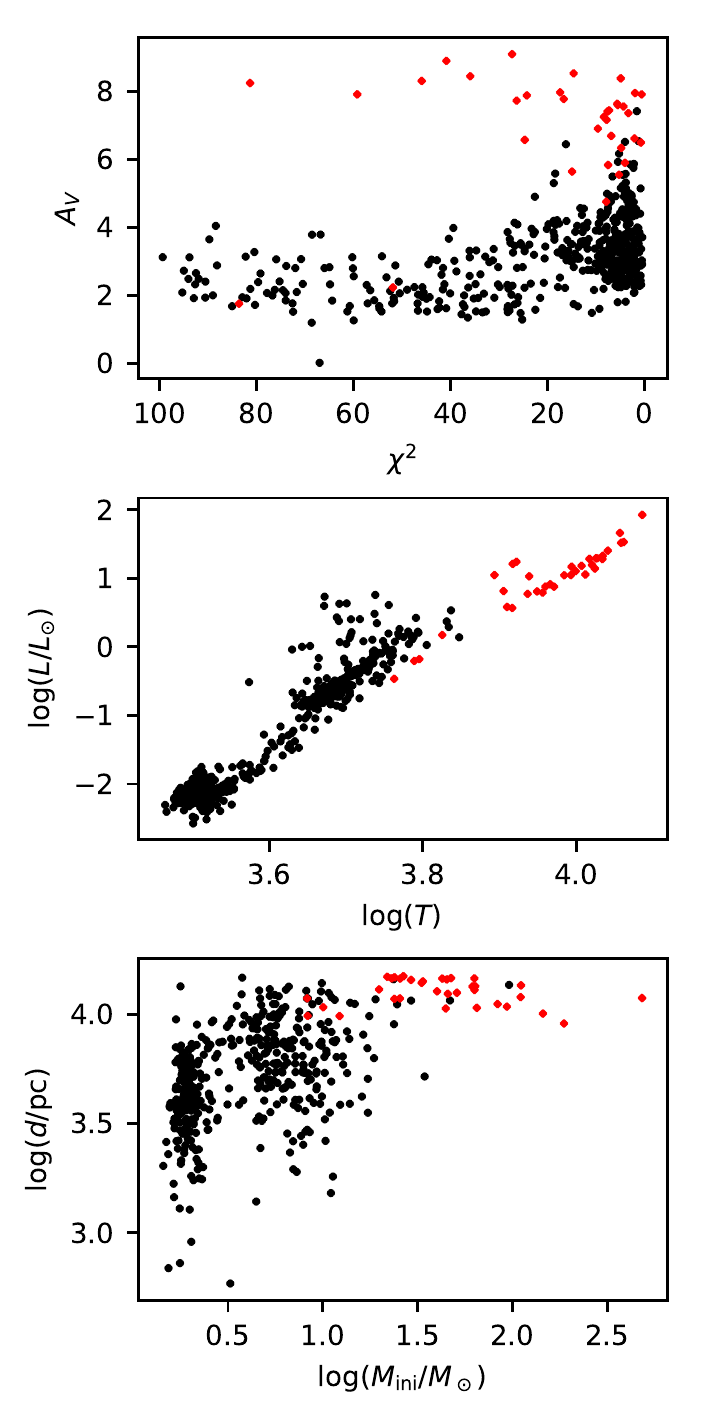}
    \caption{\edit1{Illustration of the suspicious, high-$A_V$ fit results described in section \ref{sub:postprocess}.
        Every dot corresponds to the fit result for a single star.
        \textit{Red:}} Stars removed by applying cuts on the temperature ($\log T_\text{exp}$ and $\log T_\text{best} < 3.87$).
      \edit1{Most of these high temperature stars also have high luminosities (middle panel), as well as high distances and masses (bottom panel).}
      This is suspicious, and despite their seemingly normal $\chi^2$ values (top panel), these fit results are likely to be incorrect.}
    \label{fig:agb}
\end{figure}

There also seems to be group of stars for which the $\chi^2$ is not necessarily large, yet their
$A_V$ is unusually high ($\gtrsim 7$ mag). When inspecting the other parameters for these stars (Figure \ref{fig:agb}),
they all seem to have high temperatures ($\log T \gtrsim 3.8$), luminosities
($\log L/L_{\astrosun} \gtrsim 1$), masses ($\log M_{\text{ini}} \gtrsim 1.3)$, and distances
($d \gtrsim 10000~\text{pc}$). This likely points to a type of star which the BEAST fails to fit
correctly, due to the necessary stellar models not being present. The resulting parameters might
provide an acceptable SED, but their values are likely unphysical. In the second panel of Figure \ref{fig:agb}, a gap can be seen at a temperature of $\log T = 3.87$. We choose to place a cut on the temperature there; all stars that have an expectation value or best fit value of $\log T > 3.87$ are deemed suspicious, and are removed from the catalog.
\edit1{The stars that were removed using this cut are highlighted in gray in Figure \ref{fig:corner}.}

This
phenomenon has some similarities with a previously documented failure case for the BEAST, which
occurs when an attempt is made to fit the SED model to a Thermally Pulsing Asymptotic Giant
Branch (TP-AGB) star\footnote{\url{https://beast.readthedocs.io/en/latest/beast_issues.html}}.

\subsection{$A_V$ and $R_V$ Maps}
\begin{figure*}[htbp]
  \centering
  \plotone{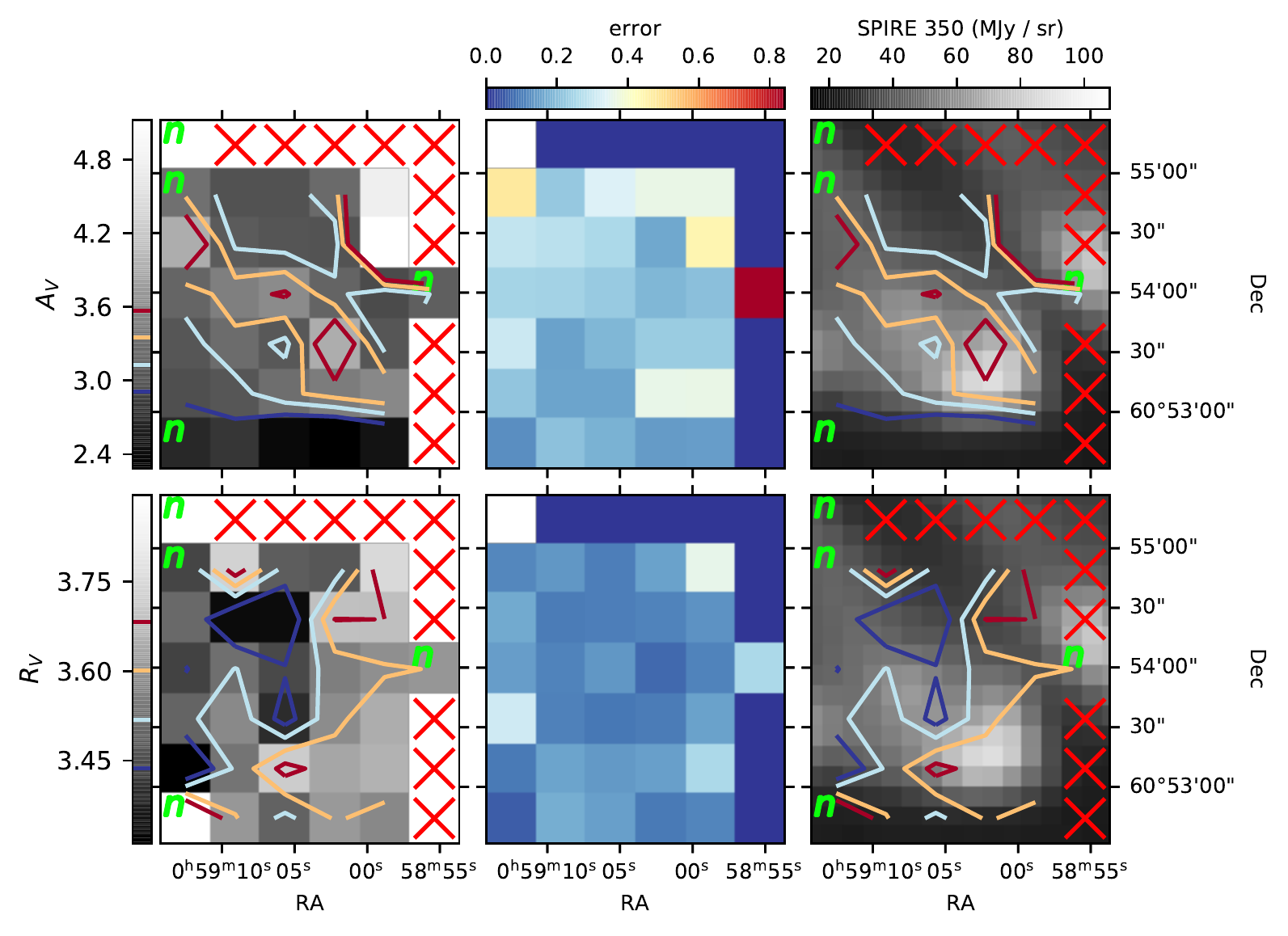}
  \caption{\edit2{\textit{Left:}} \edit1{Maps of $A_V$ and $R_V$ and their contours.
      \edit2{The contour levels are indicated on the respective colorbars.}}
    \edit1{\textit{Middle:} Errors on the map values.}
    \edit2{\textit{Right:}} The same contours overplotted onto a SPIRE 350 micron image.
    \textit{Red $X$}: No sources. \textit{Green $n$}: Low ($\leq 5$) number of sources.}
  \label{fig:maps}

  \plotone{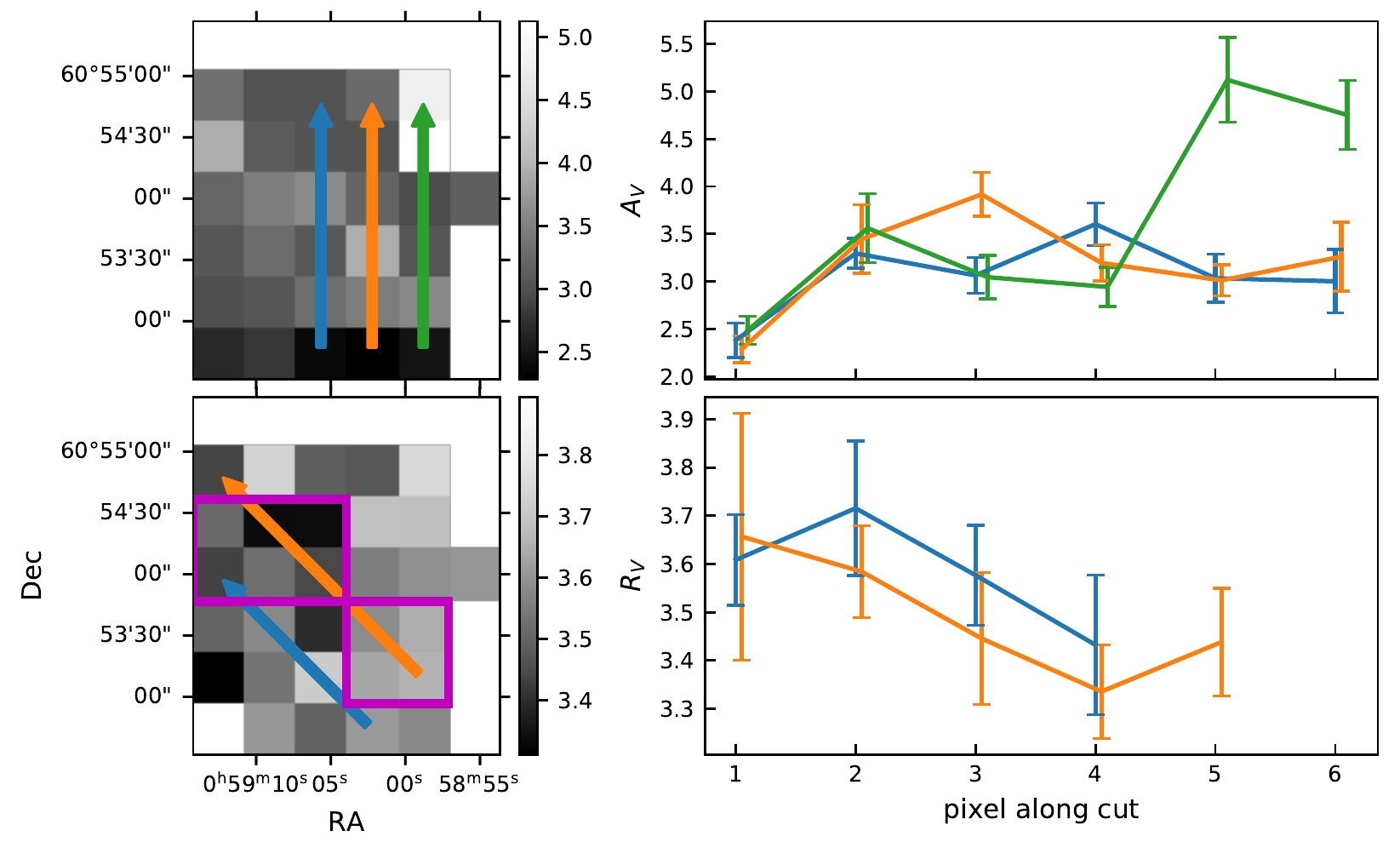}
  \caption{\edit1{Cuts through the $A_V$ and $R_V$ maps.
      \textit{Arrows:} Starting point and direction of each cut.
      \textit{Curves and error bars:} Value and error along each arrow, color coded.
      Different offsets have been added on the x-axis for legibility reasons.
      The magenta rectangles indicate the two regions used to determine the significance of the $R_V$ decrease.}}
  \label{fig:cuts}
\end{figure*}

The main objective of this work is to look for signs of dust evolution, by mapping the spatial variations of the $A_V$ and $R_V$ parameters. The maps, shown in Figure \ref{fig:maps}, are constructed
by creating a rectangular grid, aligned with the right ascension and declination axes.
The angular size of each map pixel is roughly 25 arcsec, which provides us with about 10 to
20 sources per \edit1{pixel} (except at the edges, as indicated with a green $n$ on the figure).
The $A_V$ and $R_V$ values \edit1{for these pixels} are calculated by considering all the \edit1{stars} that fall within their
boundaries, and taking the median of the expectation values.
Some of the \edit1{pixels} have no sources
in them at all, mainly due to geometry effects, as the \edit1{grid} does not line up with the
edges of the IR data.
These \edit1{pixels} are indicated with a red X.
The error for
each \edit1{pixel} is estimated by computing the standard deviation for the same set of values, and
dividing it by the square root of the number of sources $N$.
The typical error is $\sim 0.25$ mag for $A_V$, and $\sim 0.15$ for $R_V$.
\edit2{To show the size of the errors versus the size of the variations, cuts through these maps are shown in Figure \ref{fig:cuts}.}
\edit1{The numeric values of the maps are given in Table \ref{tab:avnumbers} and Table \ref{tab:rvnumbers} in the appendix.}

To highlight the variations of $A_V$ ($R_V$)
across the nebula, some contours are shown which have been calculated directly from
the maps. The contour levels for the maps were chosen by taking 4 linearly spaced values between
the 16th and 84th percentiles of the total range of $A_V$ ($R_V$) values. The same contours are
plotted onto the SPIRE 350 micron image, to indicate the spatial correspondence between features
of the $A_V$ and $R_V$ maps and those of the FIR emission.

It should be noted that $A_V$ does not drop to zero outside of the visible gas of the nebula.
In the area at the bottom of the map, just outside the visible edge of the gas, an $A_V > 2$ is still observed while the 350 micron emission reaches a minimum (near the background level of $\sim 22$ MJy / sr). A selection of 50 stars in this area, has an $A_V$ distribution centered at $\sim$ 2.5 mag, with a spread of 0.70 mag. They have a mean $A_V$ of \num{2.51 \pm 0.10}, and we assume that this is the extinction produced by material in the background (or foreground) of IC\,63.

\begin{figure}[htbp]
    \centering
    \plotone{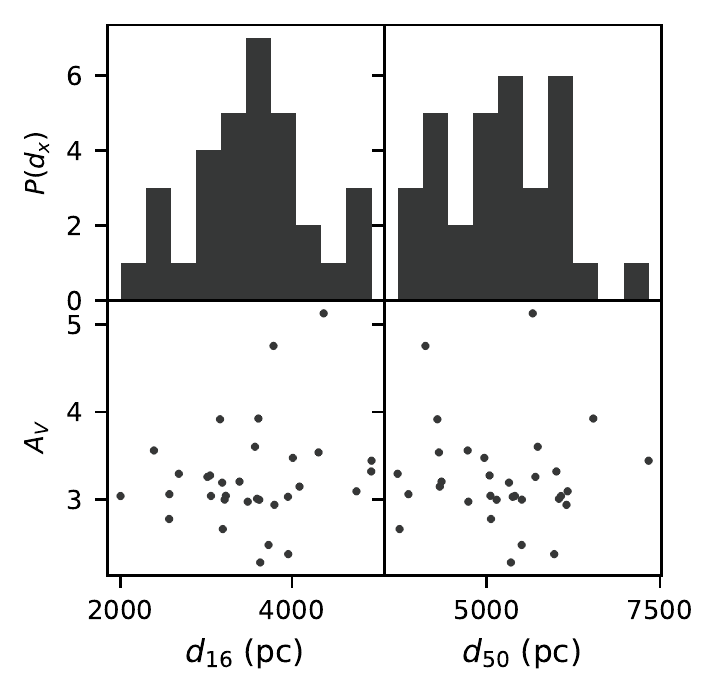}
    \caption{\edit1{Histogram and scatter plot for the distance percentiles in each pixel. Each point corresponds to one of the pixels of the $A_V$ map. The plot for $d_{84}$ is not shown here, but shows a similar scatter.}}
    \label{fig:distance}
\end{figure}

\edit1{This simplistic method of accounting for the diffuse Galactic extinction comes with an important caveat, especially since this value is larger than the residual $A_V$, i.e.~the \edit2{extinction} by the nebula.
  Since the stars are at widely varying distances (1 to 15 kpc), the extinction between each star and the cloud could be very different.
  To investigate whether this could contaminate the structure of the $A_V$ map, we examined the distance distribution in each pixel by making maps of the 16th, 50th and 84th percentiles of the observed set of distances: $d_{16}$, $d_{50}$, and $d_{84}$.
  These maps did not show any obvious structure which could drastically contaminate the observed trends in $A_V$.
  This was further confirmed by examining scatter plots (Figure \ref{fig:distance}) of the $A_V$ pixels versus the corresponding distance distribution percentiles.
  Furthermore, the range of $d_{50}$ values for the pixels is quite constrained, with almost all the values being between 4 and 6 kpc, with more or less a flat distribution.
  Similarly $d_{16}$ lies between 2000 and 4000 for most pixels, and $d_{84}$ lies between 5000 and 8000.
  Therefore, the diffuse Galactic extinction introduces an extra source of noise caused by variations in the average distance, but we do not expect it to produce false trends.}
  
\section{Discussion} \label{sec:discussion}
\subsection{$A_V$}
The $A_V$ map shows that the extinction peaks at \edit1{$00 \text{h} 59 \text{m} 02 \text{s}; \ang{60;53;30}$, with $A_V = (3.91 \pm 0.23)\,\text{mag}$.
  This is near the tip of the nebula.}
Given the background $A_V$ estimate of the previous section, the average amount of extinction added by the tip of IC\,63 is hence
\begin{equation}
    \Delta A_{V, \text{tip}} = (1.41 \pm 0.25)\,\text{mag}
\end{equation}
The higher $A_V$ values seem to lie along a ridge-like feature containing this maximum, more or less parallel to the direction of the radiation
\edit1{(see arrow in Figure \ref{fig:footprint})}.
The \edit1{extinction} of this part in the nebula typically ranges from 1.0 to 1.4
mag.
With the exception of the bottom row of the map which we used to measure the $A_V$ background, we find a minimum $A_V$ of 3.0 mag, or a \edit1{net extinction} of at least 0.5 mag for the rest of the values.

In the top right of the map, a feature with a much higher \edit1{extinction ($A_V \sim 5$ mag)} can be seen, 2.5 mag when subtracting the background.
\edit1{The cuts in Figure \ref{fig:cuts} show that this peak is significantly more pronounced than the other features.}
On the full version of the SPIRE map, \edit1{this} feature is a bright part of a larger
structure in the FIR, which does not have optical emission.
In the NIR \edit1{(e.g.~F160W in Figure \ref{fig:collage})}, it can be seen very faintly.
We do not consider this structure part of the PDR, and we ignore its contribution in the analysis that follows.

\subsection{$R_V$}
The $R_V$ measurements over individual stars have a spread of 0.5, and a mean of $R_V = 3.556 \pm 0.001$ (the error on the mean is this small due to the number of sources in our sample, but excludes systematics). The contours in Figure \ref{fig:maps} seem to indicate a slight drop in $R_V$, going from $\sim 3.6$ at the front, to $\sim 3.4$ at the back of the PDR.
\edit1{Given the errors on the map of about 0.15, this drop is not significant when comparing individual pixels.
  \edit2{Instead, we first examine the diagonal cuts in Figure \ref{fig:cuts}.
    For both the orange and the blue curves, a slope is observed, and the difference between the second and fourth point is larger than the error bars.}
  \edit2{Lastly, to quantify the significance of the $R_V$ decrease}, we focus on the two magenta regions shown \edit2{in the same figure}.
  We gather the individual $R_V$ results for the stars in each of the two regions, and calculate the means.
  For the front region (bottom right rectangle in Figure \ref{fig:cuts}), we find $R_V^\text{front} = 3.63 \pm 0.08$, and for the back region $R_{V,\text{mean}}^{\text{back}} = 3.42 \pm 0.05$.
  A standard two-sample t-test comparing the two sets of $R_V$ values yields $t = 2.25$, with a two-tailed p-value of $0.02$,}
\edit2{indicating that the mean $R_V$ differs significantly between the front and the back region}

One also has to take into account that the observed $R_V$ is a mix of that of the nebula and the background.
Since the $A_V$ of the background ($\sim 2.5$ mag) is larger than that of the nebula (0.5 to 1.4 mag), the drop in $R_V$ of the dust in IC\,63 will be stronger than what is observed.
We can write the measured $R_V$ in terms of the extinction parameters of the two parts: $(A_V^B$, $R_V^B)$ for the background, and $(A_V^N$, $R_V^N)$ for the nebula
\begin{equation}
    R_V = \frac{A_V}{E(B - V)} = \frac{A_V}{\frac{A_V^N}{R_V^N} + \frac{A_V^B}{R_V^B}}
\end{equation}
where we use the fact that the measured $E(B - V)$ is \edit1{the} sum of the two contributions.
Isolating $R_V^N$ from this equation, we obtain
\begin{equation}
    R_V^N = \frac{A_V^N}{\frac{A_V}{R_V} - \frac{A_V^B}{R_V^B}} = \frac{A_V - A_V^B}{\frac{A_V}{R_V} - \frac{A_V^B}{R_V^B}} \label{eq:rvcorrection}
\end{equation}
Knowing $A_V$ and $R_V$ from our maps, and using our value of $A_V^B = 2.51$ mag obtained earlier, we just need to assume a value for $R_V^B$. It should be noted however that equation \ref{eq:rvcorrection} is quite sensitive to the assumed value of $R_V^B$. For example, picking a value near the Galactic average of $\sim 3.0$ often leads to very high or even negative values for $R_V^N$.
If we take the average $R_V$ over the same region that we used to determine $A_V^B$, we obtain $R_V^B = 3.6$.

The galactic $R_V$ map by \citet{2017ApJ...838...36S} shows values between 3.0 and 3.7, with variations at kiloparsec scales.
While there might still be variations at smaller scales, we assume that $A_V^B$ and $R_V^B$ stay constant over our field of view.
With these assumptions, we find that $R_V^N$ ranges from 3.7 at the bottom right of the map (tip of the nebula), to values between 2.5 and 3.3 near the top left.
Since $R_V$ correlates with the average grain size, this points to the existence of processes making the average grain size of the dust population larger at the front of the nebula.
Possible candidates are coagulation and accretion at the tip of the nebula due to higher gas densities \citep{2003A&A...398..551S, 2015A&A...579A..15K}.

\citet{2013ApJ...775...84A} determined $A_V$ and $R_V$ for 14 observed background stars of
IC\,63. This was done using a spectral classification of the stars, followed by a calculation of
$R_V$ using the relation $R_V = 1.12 \times E(V - K) / E(B - V) + 0.02$
\citep{1999PASP..111...63F}. Via the total-to-selective extinction and
the color excess $E(B - V)$, $A_V$ was then obtained.
Unfortunately these stars are not present in our sample, as they were either too bright (and removed per section \ref{subsub:culling}), or simply outside of the field of view. They obtained an average value $\langle R_V \rangle = \num{2.2 \pm 0.5}$, which is significantly lower than the values listed by our $R_V$ map.
\edit1{We do not believe that this is due to different physical conditions, as the positions of the stars used in \citet{2013ApJ...775...84A} probe different areas of the cloud: some are at the front edge, and others are behind or inside the cloud.
The low $R_V$ values listed for all of these stars, and the fact that there is no overlap between the samples, make it hard to compare these results.}

\subsection{Column Density to $A_V$ Ratio}
Our results have an average $A_V$ across IC\,63 of 3.3 mag, or 0.8 mag with the background subtracted.
As a check on the validity of this measurement, we make an estimate of the $N_\text{H}$ \edit1{(column density of hydrogen nuclei)} to $A_V$ ratio.

The relationship between the extinction $A_V$ and the
column density $N_\text{H}$ of hydrogen has been found to be linear, on average in the Milky Way. Different values for the ratio
$\left\langle N_\text{H} / A_V \right\rangle$ have been found, by using far-UV extinction observations \citep{1978ApJ...224..132B, 1994ApJ...427..274D} and observations of X-ray sources
\citep{1973A&A....26..257R, 1975ApJ...198...95G, 1995A&A...293..889P, 2009MNRAS.400.2050G}.
The most recent measurements by \citet{2017MNRAS.471.3494Z} show that
$\left\langle A_V / N_\text{H} \right\rangle$ is more or less invariant across the Galaxy, and
$N_\text{H} = \num{2.08 \pm 0.02 e21} A_V \, \si{\per\square\cm}\,\text{mag}^{-1}$.

The HI4PI 21 cm survey \citep{2016A&A...594A.116H} \edit1{contains two sightlines which cover our area of interest;
  the listed H\textsc{i} column densities are \SI{4.47e21}{\per\square\cm} and \SI{5.53e21}{\per\square\cm}.
  We assume that the average atomic hydrogen column density for IC\,63 is somewhere between these two values:}
$N_\text{H\textsc{i}} = \SI{4.50 \pm 0.03 e21}{\per\square\cm}$.
Since we can not distinguish the 21 cm contribution of IC\,63 from that of the background, we assume that the $N_\text{H\textsc{i}} / A_V$ ratio is the same for IC\,63 and its background, and we divide by \edit1{the average} $A_V^\text{total} = 3.3~\text{mag}$.
\begin{equation}
\frac{N_\text{H\textsc{i}}}{A_V} = \frac{N_\text{H\textsc{i}}^\text{total}}{A_V^\text{total}} = \SI{1.36e21}{\per\square\cm}\,\text{mag}^{-1}
\end{equation}

\citet{2018A&A...619A.170A} used H\textsubscript{2} excitation diagrams based on IRS spectra
from \textit{Spitzer} to determine the gas temperature and column density for the molecular component. They found that it consists of two components, of which the cooler one has a temperature
of $T = \SI{207 \pm 30}{\kelvin}$ and a column density of
\edit1{$N_{\text{H}_2} = \SI{2.3e20}{\per\square\cm}\text{,}$}
while the warm component has
$T = \SI{740 \pm 47}{\kelvin}$ and $N_{\text{H}_2} = \SI{9.3e17}{\per\square\cm}$.
Given the dominant (cold component) value from \citet{2018A&A...619A.170A} for the column
density of H$_2$, and our background subtracted value for the average $A_V = 0.8~\text{mag}$, we find a $2N_{\text{H}_2} / A_V$ ratio of
\begin{equation}
  \frac{2N_{\text{H}_2}}{A_V} = \SI{5.75e20}{\per\square\cm} \text{mag}^{-1}
\end{equation}

\edit1{Since we are looking at the PDR, we expect the medium to be mostly neutral. IPHAS data \citep{2005MNRAS.362..753D, 2014MNRAS.444.3230B} show that most of the H$\alpha$ emission is emitted at the bright, irradiated edge of the cloud.
\edit2{Additionally}, the PDR model of \citet{1995A&A...302..223J} predicts that the H$^+$ abundance is at least 6 orders of magnitude lower than that of H or H$_2$.
Therefore, we will not take into account the column density of ionized hydrogen here.}
The combined value is
\begin{equation}
    \frac{N_\text{H}}{A_V} = \frac{N_\text{H\textsc{i}}}{A_V} + \frac{2N_{\text{H}_2}}{A_V} = \SI{1.94e21}{\per\square\cm}\,\text{mag}^{-1}
\end{equation}
This value is quite close to the value from the literature mentioned above.
Looking for a separate velocity component in the H\textsc{i} data, and combining our measurements with better MIR spectroscopy, would provide a way to \edit1{map} the molecular fraction and the gas to dust ratio for IC\,63.

\subsection{$A_V$ to Surface Brightness Relation}
\begin{figure*}
  \centering
  \plotone{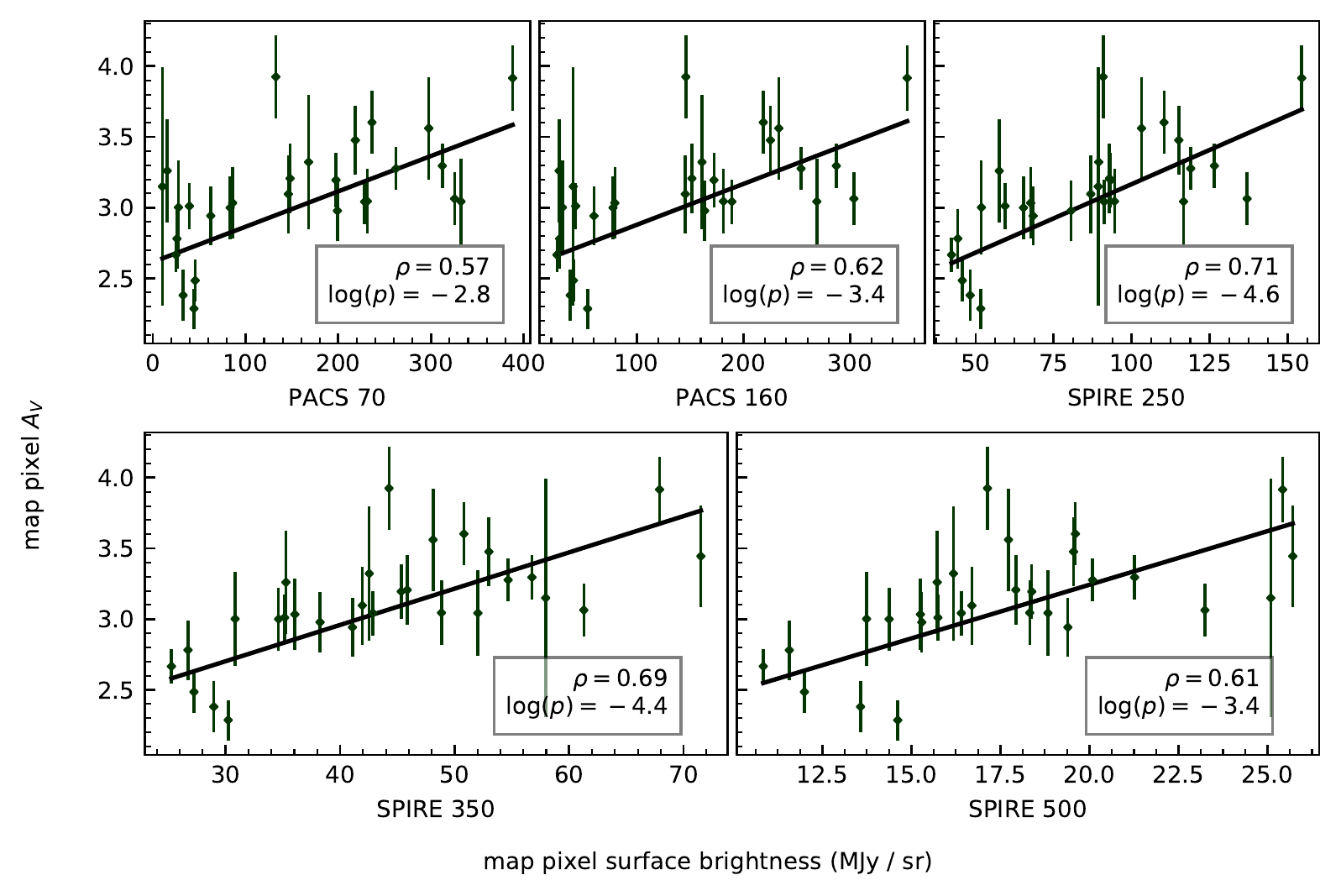}
  \caption{Correlation of $A_V$ with the reprojected \textit{Herschel} fluxes. The
    black line shows the best fit slope. \edit1{The Pearson correlation coefficient $\rho$ is shown, together with} the p-value for a two sided hypothesis test, where a slope of 0 is the null hypothesis.}
  \label{fig:herschelscatter}
\end{figure*}

\begin{figure*}
  \centering
  \plotone{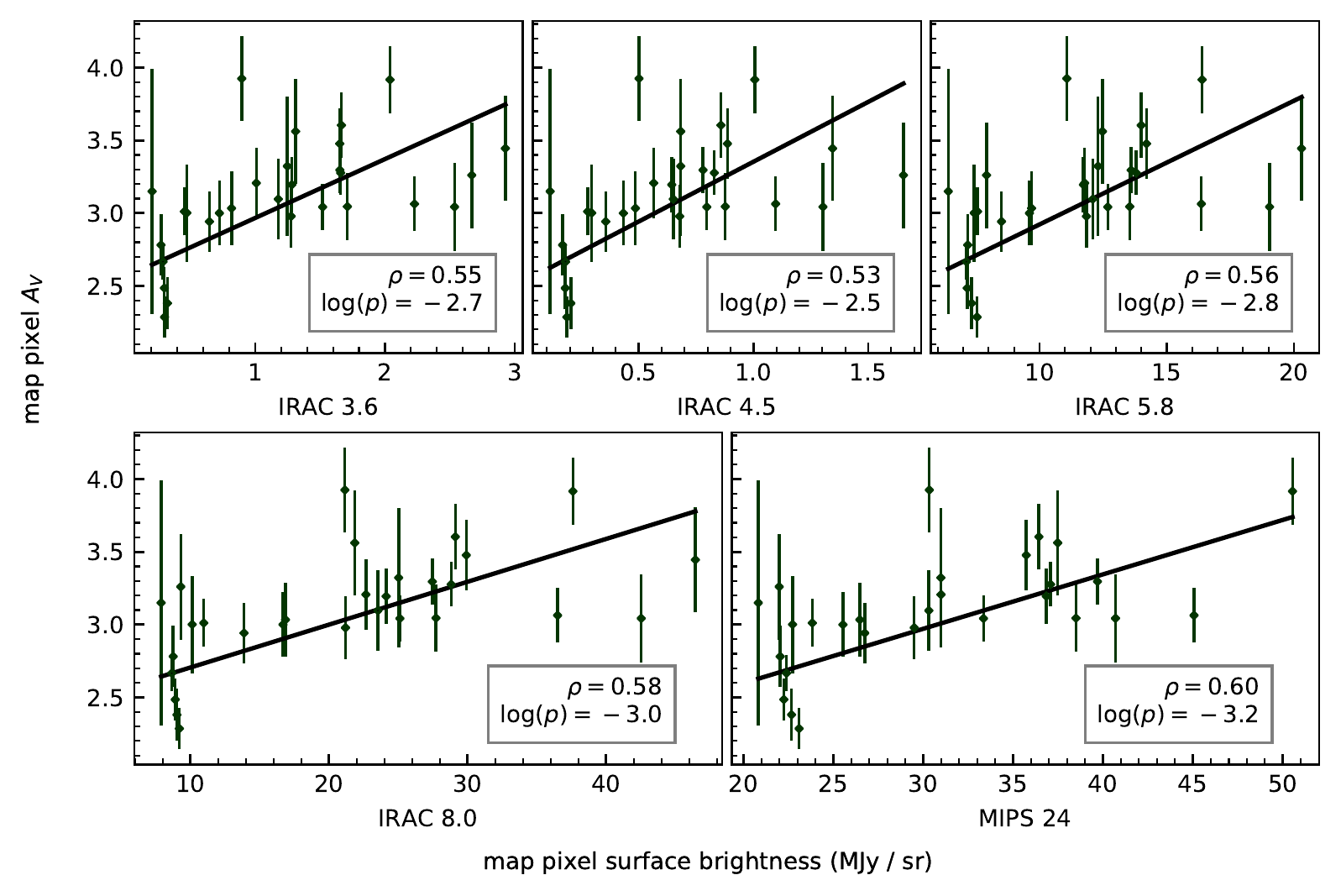}
  \caption{Correlation of $A_V$ with the reprojected \textit{Spitzer} fluxes, analogous to Figure \ref{fig:herschelscatter}}
  \label{fig:spitzerscatter}
\end{figure*}

We investigate how our $A_V$ and $R_V$ maps compare to the dust emission measured using \textit{Spitzer} and \textit{Herschel} data. To look for a relation between $A_V$ and the surface brightness in the different bands, the IRAC, MIPS, PACS and SPIRE maps are first reprojected onto the same coordinate grid, using the \textit{reproject} package
for Python.\footnote{\url{https://reproject.readthedocs.io/en/stable/}}
For each reprojected image, the pixels are matched with the ones from the $A_V$ map, and the reprojected
fluxes are plotted against the $A_V$ map values, as shown in Figures \ref{fig:herschelscatter} and \ref{fig:spitzerscatter}.
\edit1{For all bands, we find a positive correlation between $A_V$ and the observed flux.}
Upon applying the same analysis to the $R_V$ map, no significant correlation between any of the fluxes and $R_V$ was found.

\begin{figure*}
  \centering
  \plotone{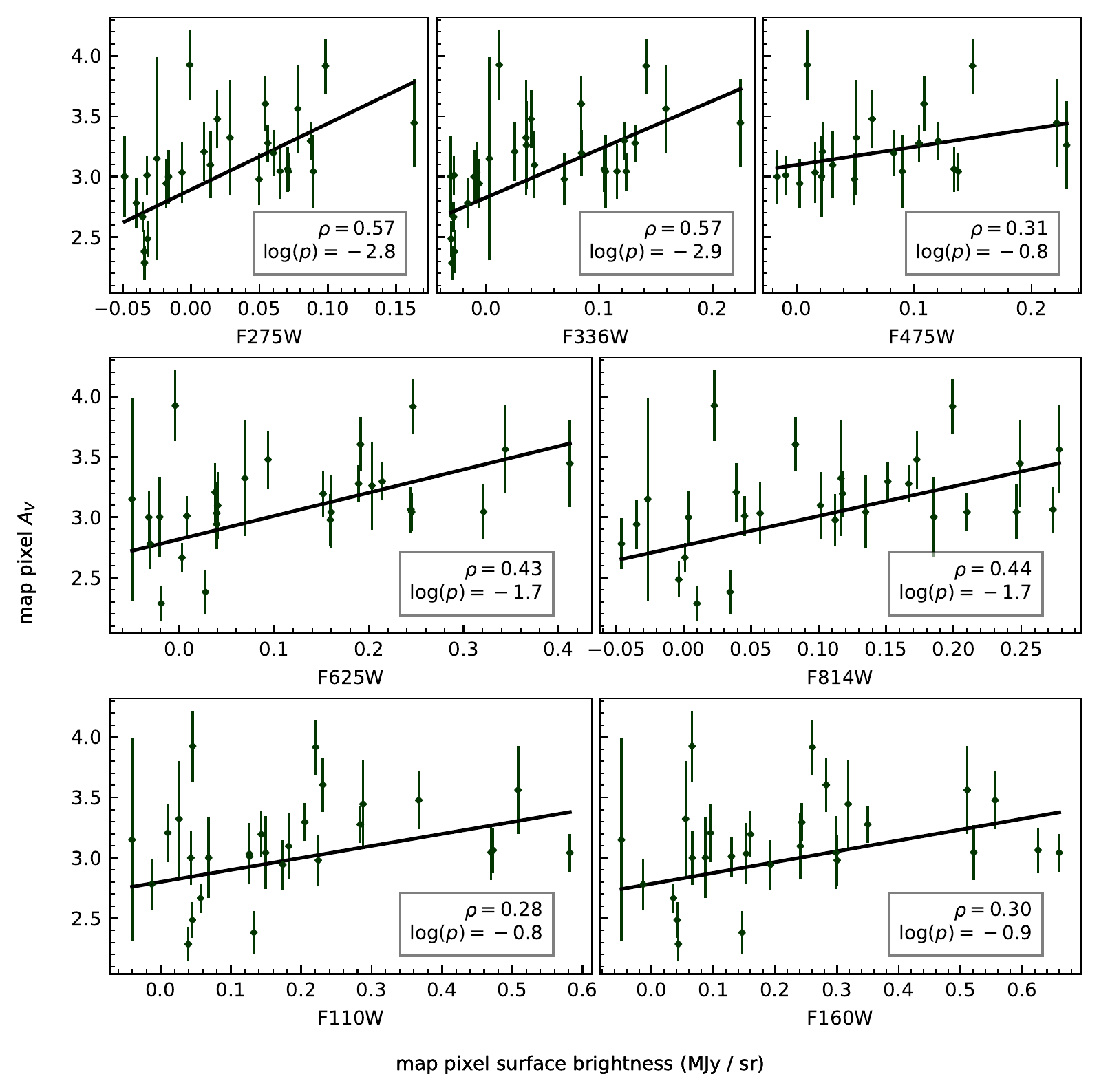}
  \caption{Analogous to Figures \ref{fig:herschelscatter} and               \ref{fig:spitzerscatter}, for the HST data. Some of the correlations are somewhat less significant, as can be seen from the $\rho$ and $p$ values, but we still obtain usable values for the slopes, albeit with larger error bars.}
  \label{fig:hstscatter}
\end{figure*}

\edit1{For each panel in Figures \ref{fig:herschelscatter} and \ref{fig:spitzerscatter},
we determine a slope and an uncertainty using a linear fit, which is weighted according to the errors on the $A_V$ pixels.
The inverse of this slope gives us the the flux produced per $A_V$-unit of material present, which can be interpreted as an emission coefficient for the dust.
Figure \ref{fig:slopes} shows these inverse slopes as a function of wavelength.}
\edit1{We applied the same technique to our HST data (Figure \ref{fig:hstscatter}), showing that this relation extends into the visual and UV, albeit less significant and with greater scatter.
  As a consequence, the error bars for the emission coefficients are larger.
This is to be expected, as the observed flux is more complex for these wavelengths.
There are contributions from the gas emission, as well as scattered light from $\gamma$ Cas.}

\edit1{These measurements can serve as useful observational constraints for future SED modeling efforts of IC\,63 or similar objects, especially those that model the dust composition and emissivity.
We therefore present the values and their uncertainties in Table \ref{tab:slopes}. In the next section, we will use these results to fit a simple dust emission model.}

\begin{deluxetable}{lrr}
\tablecaption{Measurements and uncertainties of the average surface brightness per $A_V$ unit. \label{tab:slopes}}
\tablehead{\colhead{Band} & \colhead{$\Delta S_\nu(\lambda) / \Delta A_V$} & \colhead{$\sigma$}\\
           \colhead{} & \colhead{($\text{MJy}\,\text{sr}^{-1}\,\text{mag}^{-1}$)} & \colhead{($\text{MJy}\,\text{sr}^{-1}\,\text{mag}^{-1}$)}}
      \startdata
F275W & 0.183 & 0.025 \\
F336W & 0.250 & 0.035 \\
F475W & 0.669 & 0.350 \\
F625W & 0.520 & 0.094 \\
F814W & 0.408 & 0.069 \\
F110W & 1.007 & 0.239 \\
F160W & 1.115 & 0.242 \\
IRAC 3.6 & 2.472 & 0.332 \\
IRAC 4.5 & 1.216 & 0.164 \\
IRAC 5.8 & 11.770 & 1.563 \\
IRAC 8.0 & 33.961 & 4.387 \\
MIPS 24 & 26.741 & 3.496 \\
PACS 70 & 399.678 & 54.112 \\
PACS 160 & 345.684 & 45.670 \\
SPIRE 250 & 103.251 & 12.661 \\
SPIRE 350 & 38.976 & 4.734 \\
SPIRE 500 & 13.184 & 1.731 \enddata
\end{deluxetable}

\subsection{Model for the Average Dust SED}
\begin{figure}
  \centering
  \plotone{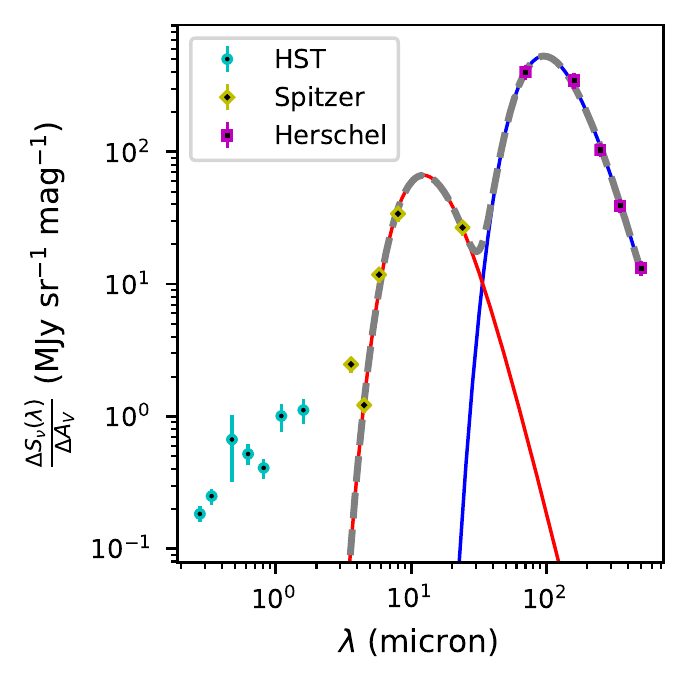}
  \caption{Surface brightness per $A_V$ unit for each of the MIR and FIR observations
    provided by \textit{Spitzer} and \textit{Herschel}, and modified blackbody fits to these
    data. The points for the HST data are added for completeness, but are not used in the blackbody fits. The data points are calculated by fitting slopes to the scatter plots in Figures \ref{fig:herschelscatter}, \ref{fig:spitzerscatter} and \ref{fig:hstscatter}. \textit{Red:} Fit to the
    IRAC and MIPS 24 points, ignoring the IRAC 3.6 point.
    \textit{Blue:} Fit to the PACS and SPIRE points.
    \textit{Gray dashed:} Combined model, fitting both sets of points.}
  \label{fig:slopes}
\end{figure}

The surface brightness of dust with a single equilibrium
temperature $T_d$ is given by
\begin{equation}
  S_\nu(\lambda) = \kappa(\lambda) \Sigma_d B_\nu(\lambda; T_d)
\end{equation}
where $\kappa$ is the grain absorption cross section per unit dust mass as a function of
$\lambda$, $\Sigma_d$ is the dust surface mass density and $B_\nu$ is the Planck function. The
subscript $\nu$ indicates that we are using quantities per unit frequency (e.g. MJy). Within
this model with $\kappa$ and $\Sigma_d$ as parameters, $\kappa$ and $\Sigma_d$ are degenerate. But the quantity that we have measured, that is displayed in Figure
\ref{fig:slopes} is actually the surface brightness per unit of $A_V$:
\begin{equation}
  \frac{S_\nu(\lambda)}{A_V} = \kappa(\lambda) \frac{\Sigma_d}{A_V} B_\nu(\lambda; T_d)
\end{equation}
So if a value for the ratio $\Sigma_d / A_V$ is known, it becomes possible to
measure $\kappa$. In practice, we use a modified blackbody model of the form
\begin{equation}
  \label{eq:modbb}
  \frac{S_\nu(\lambda)}{A_V} = \kappa_{\text{eff}, 160} \frac{\Sigma_d}{A_V} \left( \frac{\lambda}{\SI{160}{\micro\m}} \right)^{-\beta_{\text{eff}}} B_\nu(\lambda; T_{\text{eff}})
\end{equation}
as in \citet{2014ApJ...797...85G}. \edit1{The scale factor $\kappa_{\text{eff}, 160}$} determines the effective value of $\kappa$ at 160 micron.
The second and third
parameters are the \edit1{effective spectral index} $\beta_{\text{eff}}$ and the effective
temperature $T_{\text{eff}}$.
These parameters are called effective because the observed surface brightness per
$A_V$ is a mix of measurements along different sightlines, and each sightline is a mix of
different dust populations.
By using the well known relationship
\begin{equation}
  A_V = 1.086\,\tau_V\,\text{mag} = 1.086 \kappa_{\text{eff}, V} \Sigma_d\,\text{mag} \label{eq:avtau}
\end{equation}
we can write the model as
\begin{align}
    \frac{S_\nu(\lambda)}{A_V} &= \frac{\tau_{160}}{A_V} \left( \frac{\lambda}{\SI{160}{\micro\m}} \right)^{-\beta} B_\nu(\lambda; T) \\
    &= \frac{\kappa_{160}}{1.086 \kappa_V} \left( \frac{\lambda}{\SI{160}{\micro\m}} \right)^{-\beta} B_\nu(\lambda; T)\,\text{mag}^{-1}
\end{align}
where we drop the `$\text{eff}$' subscript for conciseness. Therefore, by fitting this model to our measurements we can determine the average \SI{160}{\micro\metre} to V band opacity ratio. We treat the ratio $\tau_{160} / A_V = \kappa_{160}/ 1.086 \kappa_V\,\text{mag}^{-1}$ as a single fit parameter.

We found that the sum of two modified blackbody models works well to fit the slopes calculated
from the \textit{Herschel} and \textit{Spitzer} maps simultaneously. We use a least squares fitting approach using the standard options of
Astropy's \texttt{modeling} module. Three fit results are displayed in Table \ref{tab:mbb}, and the corresponding curves are shown in \edit1{Figure} \ref{fig:slopes}.

\begin{deluxetable*}{rcccccc}
  \tablecaption{Overview of the modified blackbody fits to our measurement of $\Delta S_\nu(\lambda) / \Delta A_V$. \label{tab:mbb}}
  \tablehead{\colhead{Fitting method} & \colhead{$\tau_{160; 1} / A_V$} & \colhead{$\beta_1$} & \colhead{$T_1$} & \colhead{$\tau_{160; 2} / A_V$} & \colhead{$\beta_2$} & \colhead{$T_2$}\\
  \colhead{} & \colhead{($10^{-4} \text{mag}^{-1}$)} & \colhead{} & \colhead{(\si{\kelvin})} & \colhead{($10^{-9} \text{mag}^{-1}$)} & \colhead{} & \colhead{(\si{\kelvin})}}
  \startdata
  Individual\tablenotemark{a}, free $\beta$ & \num{6.4 \pm 1.9} & \num{2.06 \pm 0.22} & \num{29.9 \pm 2.2} & \num{1.4 \pm 2.4}  & \num{2.4 \pm 1.0} & \num{227 \pm 6}\\
  Individual, fixed $\beta$ & \num{6.43 \pm 0.40} & 2.06 & \num{29.9 \pm 0.5} & \num{2.70 \pm 0.71} & 2.06  & \num{227 \pm 6}\\
  Simultaneous\tablenotemark{b}, fixed $\beta$ & \num{6.43 \pm 0.81} & 2.06 & \num{29.9 \pm 1.1} & \num{2.68 \pm 0.49} & 2.06 & \num{227 \pm 4.2}
  \enddata
  \tablenotetext{a}{Model 1 is fit to the \textit{Herschel} points, model 2 to the \textit{Spitzer points}.}
  \tablenotetext{b}{The two models and sets of points are combined.}
\end{deluxetable*}

The first modified blackbody was fitted to only the PACS and SPIRE points, with a total $\chi^2$ of 0.61.
The second model fit included the IRAC points and the MIPS 24 micron point, but not the IRAC 3.6 micron point.
\edit1{The latter was excluded because it is systematically higher than the IRAC 4.5 point, preventing the model from properly fitting the other data points.}
For this
model, the $\kappa$-ratio and $\beta$ parameters are not well constrained, and $\chi^2$ in this case is 3.55.
The estimated Pearson correlation coefficients between the three parameters are all over 90\%, which is a known problem for this parametrization \edit1{\citep{2014ApJ...797...85G}}.

Fixing $\beta = 2.06$ for both models
gives stronger constraints on the other two parameters, especially on the temperatures.
Lastly, we simultaneously fitted the sum of the two modified blackbodies with the same fixed value
for $\beta$, leading to the gray, dashed curve in Figure \ref{fig:slopes}.
For this two-component
modified blackbody model, $\chi^2$ is 4.18 when fitting the 4 free parameters to our 9 data points, and the resulting parameters are shown on the third line of Table \ref{tab:mbb}.

\edit1{Before we discuss these fit results, some remarks should be made about the meaning of these two models and their temperatures.
  The use of two modified blackbody components does not relate to the existence of a warm and a cool component of the gas in the cloud \citep{2018A&A...619A.170A}.
  Instead, they each model a different wavelength range in the SED, for which the observed flux is emitted by a different subpopulation of the grains.}

\edit1{The first component and its temperature parameter $T_1$ can be interpreted rather straightforwardly, as the observed flux at the \textit{Herschel} wavelengths consists of thermal continuum emission from big grains.
  These grains are in thermal equilibrium with the local radiation field, and $T_1$ can be used as a measure of their average temperature.
  This temperature is expected to be higher for grains residing in the low density PDR region where the radiation field is stronger, and lower for those in the cold dense parts of the cloud.}

\edit1{On the contrary, the flux in the \textit{Spitzer} bands originates from very small grains (VSGs) and polycyclic aromatic hydrocarbons (PAHs).
  The VSGs are heated stochastically by photons, and the resulting SED shows a complex \edit2{behavior} with several peaks in the MIR \citep{2001ApJ...551..807D}.
  \edit2{Despite} this, the MIR SED (except the IRAC 3.6 micron point) seems to be well described by a modified blackbody with effective temperature $T_2$.}

Using the same PACS 70 and 160 $\mu$m images, and assuming a fixed $\beta = 1.8$, \citet{2018A&A...619A.170A} derive a color dust temperature of 30 K in IC\,63. They also perform single-component modified blackbody fits with the same fixed $\beta$ at the tip of the nebula, and again obtain a dust temperature of $\sim30$ K. This matches well with our value for $T_1$.
For the gas temperature, they report two values: $207 \pm 30$ K from H\textsubscript{2} excitation diagrams, and 130 K from a PDR model.
\edit1{Note that incidentally, these values for the temperature are of the same order of magnitude as $T_2$.
  We find it unlikely however that these two values would be coupled, as the physics driving the MIR emission by VSGs and PAHs is very different from the processes driving the $H_2$ excitation \citep{2011A&A...527A.122H}.}

As for the optical depth, \citet{2018A&A...619A.170A} find that $\tau_{160} = \num{8e-4}$.
\edit1{Using} our value of $A_{V, \text{tip}} = 1.41\,\text{mag}$ \edit1{yields} $\tau_{160} / A_V = \num{5.7e-4}\,\text{mag}^{-1}$, which is compatible with our value for $\tau_{160; 1} / A_V$.
Multiplying the last two values for $\tau_{160} / A_V$ in \edit1{Table} \ref{tab:mbb} with 1.086 (eq. \ref{eq:avtau}) gives us the result
\begin{gather}
    \frac{\kappa_{160; 1}}{\kappa_V} = \num{6.98 \pm 0.88 e-4}\\
    \frac{\kappa_{160; 2}}{\kappa_V} = \num{2.91 \pm 0.53 e-9}
\end{gather}
In the above equations $\kappa_V$ is the \edit1{combined} V-band cross section of the two components.
\edit1{Since the first component dominates, the value of $\kappa_{160; 1} / \kappa_V$ will depend on the optical properties of the big grains.
The second component makes only a small contribution to the total value, and therefore $\kappa_{160; 2} / \kappa_V$ will also depend on the relative mass density of the very small grains and PAHs.}
We can compare the value of $\kappa_{160; 1} / \kappa_V$ to theoretical dust models, using $\kappa_{0.54}$ (the cross section at \SI{540}{\nano\metre}) as an approximation for $\kappa_V$.
With their standard grain size distributions, we find $\kappa_{160} / \kappa_{0.54} = \num{1.4e-3}$ for the model of \citet{2007ApJ...657..810D}, and $\kappa_{160} / \kappa_{0.54} = \num{7.4e-4}$ for the THEMIS model \citep{2017A&A...602A..46J}.

\subsection{Per-pixel Modified Blackbody Fits}
\begin{figure*}
    \centering
    \plotone{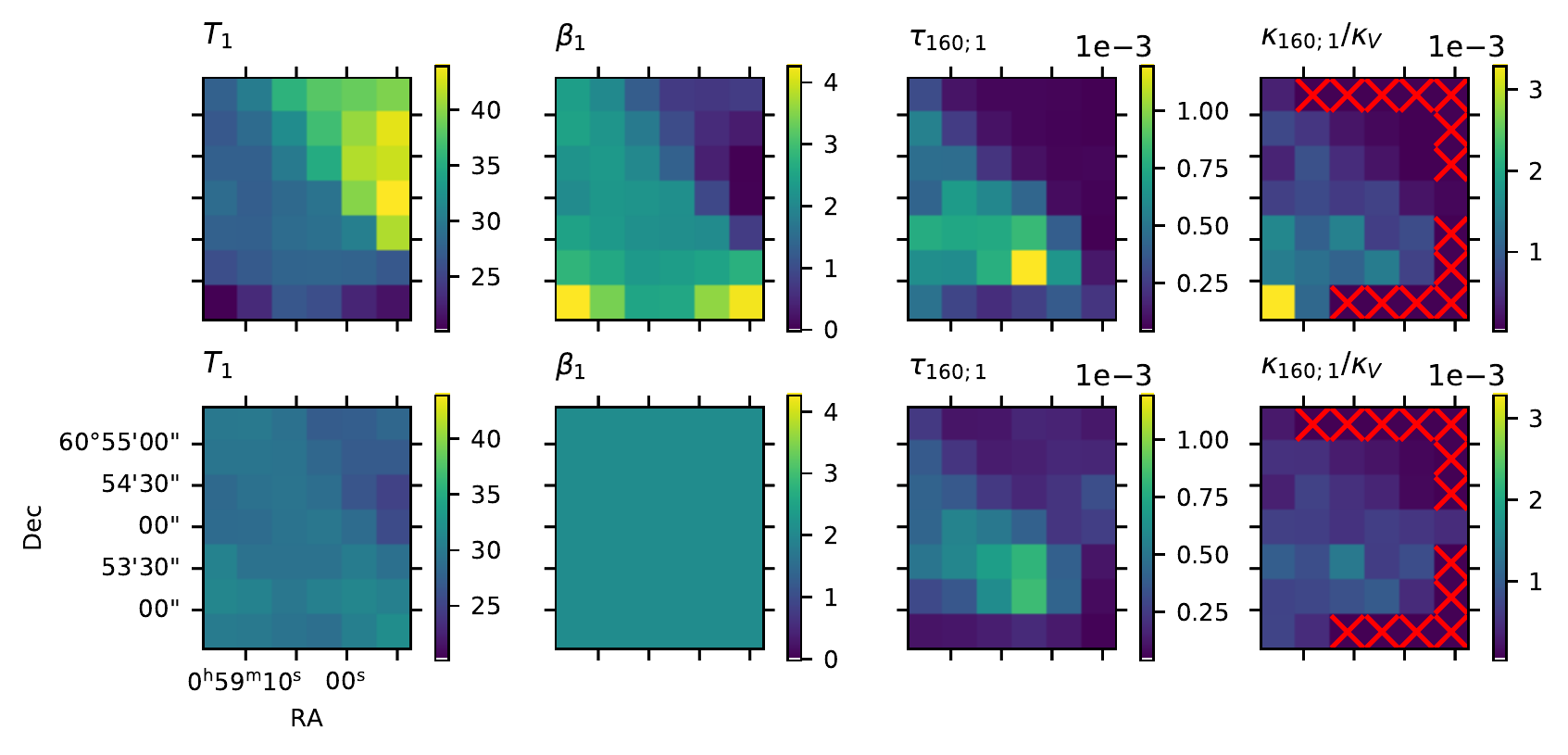}
    \caption{\edit1{Results of the per-pixel modified blackbody fitting with free $\beta_{1}$ (\textit{top row}) and fixed $\beta_1 = 2.06$ (\textit{bottom row}). \textit{Red X:} bad value. This is mainly due to missing $A_V$ or $A_V < 2.51$ (the assumed background).
        The top and bottom panel for each parameter use the exact same color scale.}}
    \label{fig:parametermaps}
\end{figure*}

Instead of making a fit to the average $\Delta S_\nu(\lambda) / \Delta A_V$ SED obtained from the slopes, we also perform individual fits for each pixel of the maps, based on the same reprojected \textit{Spitzer} and \textit{Herschel} images.
In this case, the scale parameter for the modified blackbodies is $\tau_{160}$ instead of $\tau_{160} / A_V$, \edit1{since we are fitting directly to the observed fluxes, instead of average flux versus $A_V$ slopes.}
We are not fitting a differential measurement this time, and unlike the previous section we need to determine and subtract the background in each band.
Since we have a measurement of the $A_V$ background of 2.51 mag, we use the slope and intercept of the lines in Figures \ref{fig:spitzerscatter} and \ref{fig:herschelscatter} to calculate an appropriate level for the background in each band.
By doing these per-pixel fits, we obtain maps of the same resolution for each parameter, \edit1{shown in Figure \ref{fig:parametermaps}}.
\edit1{We do not show the parameters for the MIR model, as $T_2$ and $\beta_2$ are quite noisy, and the $\tau_{160; 2}$ map looks very similar to that of $\tau_{160; 1}$, besides being of a different order of magnitude.}

Figure \ref{fig:parametermaps} shows a gradient in the temperature, perpendicular to the direction of the incoming radiation field. The typical uncertainty on $T_1$ is about 2 to 5 K.
A feature of the same shape exists for the $\beta_1$ and $\tau_{160; 1}$ parameters too, which shows an anticorrelation with the temperature. The $\beta - T$ relationship is well known and can be explained by variations in the dust properties \edit1{\citep{2003A&A...404L..11D, 2015A&A...577A.110Y}}.
However, because of noise and temperature variations along the line of sight, least-squares fitting naturally leads to this type of correlation \edit1{\citep{2009ApJ...696..676S, 2009ApJ...696.2234S, 2012A&A...541A..33J}}.
\edit1{Therefore, we performed the same fits with $\beta$ fixed to 2.06.
  A very notable difference is that the trend in $T$ becomes \edit2{inverted}, and that the temperature range becomes narrower.
  Because the $\beta - T$ trend appears partly due to noise-induced correlations and partly due to a physical relationship, these results are not easily interpreted.
  On the one hand, fixing $\beta$ might elimitate any exaggerated variations of $T$ that occur due to the correlations in the fit.
  On the other hand, choosing a constant $\beta$ value introduces a bias in $T$, which will depend on the real value of $\beta$.
  A more in-depth analysis (e.g. \citealt{2018MNRAS.476.1445G}) might be able to disentangle these degeneracies, but would go beyond the scope of this work.}

Comparing the $\tau_{160; 1}$ map to the $A_V$ map in Figure \ref{fig:maps}, we find a similar structure.
By dividing the $\tau_{160; 1}$  values by the $A_V$ value in each pixel (minus the background of 2.51 mag), and using $A_V = 1.086 \tau_V$, we obtain maps of $\kappa_{160; 1} / \kappa_V$.
\edit1{We see that this ratio is quite noisy, and comparing the top and bottom row of Figure \ref{fig:parametermaps}, fixing $\beta$ seems to affect the pixels in the bottom left corner more than others.
  The general trend is that $\kappa_{160} / \kappa_V$ seems to be higher towards the lower edge of the image.
  So while the $R_V$ map points to an increase average grain size at the front of the nebula, this analysis of the \textit{Herschel} data and the $A_V$ map shows a less pronounced trend which more or less aligns with its south edge.}

\edit1{We did not find any significant spatial correlation between $A_V$ or $R_V$ and any of the fitted parameters, except $A_V$ and $\tau$.
  This holds regardless of whether a free or fixed $\beta$ was used.
  For some of the physical conditions, \citet{2010ApJ...725..159F} provide maps based on \textit{Spitzer} IRS data:
  total PAH emission and band ratios; H\textsubscript{2} emission, column density, temperature, and ortho-to-para ratio; and the radiation field $G_0 / n_\text{H}$.
  These maps cover a 40 by 40 arcsec area at the very tip of the nebula, which corresponds to about 4 pixels of our $R_V$ map.
  Figure 7 of \citet{2010ApJ...725..159F} shows an arc shaped feature in the PAH emission, where the H\textsubscript{2} column density is also the highest.
  This feature might be related to the higher $R_V$ area observed at the front of the nebula, but the resolution of our $R_V$ measurements is insufficient confirm this.}

\subsection{Comparison with other clouds}
\edit1{In the Orion bar, the dust temperature shows a much stronger gradient, going from 71 K in front of the bar, to 49 K inside, and 37 K behind the bar \citep{2012A&A...541A..19A}.
  The distance between the first and last point for which these temperatures were measured, is of the order $0.1$ pc.
  The size of the area we study in IC\,63 is $\sim2'$ or also $\sim0.1$ pc, assuming a distance of 190 pc.
  Over the same spatial scale, there is at best a subtle temperature gradient in IC\,63.
  \edit2{Note that the radiation field and the density are more than an order of magnitude higher in the Orion bar: $G_0 = \num{4e4}$ and $n_\text{H} = \SI{2e5}{\per\cubic\cm}$.}
\edit2{Reminding the reader, the conditions in IC\,63 are $G_0 = 1100$ (150 in \citealt{2018A&A...619A.170A}), and $n_\text{H} = \SI{1.2e4}{\per\cubic\cm}$.}
  \citet{2012A&A...541A..19A} measure the values $\beta =$ 1.2, 1.6, and 2.2 for the three same points, respectively. These differences likely originate from dust evolution in the Orion bar.
  For IC\,63 we find it hard to determine the significance of the variations in the fitted $\beta$, and we observe no correlations between the changes in $\beta$ and the observed differences in $R_V$.}

\edit1{In \citet{2016ApJ...830..118S}, both the bar and the H\,\textsc{ii} region of the Orion nebula were studied using MIR observations with the FORCAST instrument on board SOFIA, and \textit{Herschel}/PACS maps.
  A two-component modified blackbody model was necessary to properly fit the thermal dust emission.
  The cool component ($\sim 40$ K) was associated with the molecular cloud, and the warm component ($\sim 80$ K) with the dust in the PDR/H\textsc{ii} region.
  For IC\,63, only one component is needed to fit the thermal dust emission; we added a second component because it was able to describe the MIR emission of the PAHs and VGSs rather well.}

\edit1{For NGC 7023, \citet{2014A&A...569A.109K} performed similar, per-pixel modified blackbody fits (with $\beta$ fixed to 2).
  The dust temperature ranges from 20.1 to 30.3 K in NGC 7023 NW, and from 17.2 to 20.5 K in NGC 7023 E.
  The temperature  differences are smaller than in the Orion bar, and closer to what we observe in IC\,63.
  Through the $\tau$ factor of their model (in this case at \SI{250}{\micro\metre}), they derive line-of-sight column densities ranging from 2.8 to \SI{9.8e22}{\per\square\cm}, more than an order of magnitude larger than in IC\,63.
  This means we can expect $A_V$ values along the line of sight of several tens of magnitudes.}
\edit2{For the radiation field, \citet{2012A&A...542A..69P} provide the values $G_0 = 2600$ for NGC 7023 NW, and $G_0 = 250$ for NGC 7023 E.
  The density $n_\text{H}$ of these two PDRs is about \num{2e4} to \SI{3e4}{\per\cubic\cm}.
  Despite the differences in radiation field \edit2{and optical depth}, the dust temperatures in NGC 7023 NW, NGC 7023 E, and IC\,63 do not show as much variation as those in the Orion bar.
  Concluding, the observed dust variations are likely related to the density $n_\text{H}$.}

\edit1{Unfortunately, we cannot compare our $R_V$ measurements to other PDRs, because direct extinction measurements in other PDRs are rarely possible, especially in a spatially resolved way.
Reiterating, the number of background stars and the low optical depth make IC\,63 one of the only PDRs for which this is possible, at least to our knowledge.}

\section{Conclusion}
We have expanded the BEAST code to better support the fitting of background stars of nebulae and other nearby, transparent objects.
We have performed individual SED fits of a combined stellar and extinction model to the
numerous background stars of the IC\,63 PDR, as observed by HST in seven broadband filters
from the UV to the NIR.
  
From these fit results, we were able to generate maps of the extinction parameters $A_V$
and $R_V$ across the nebula. We find that $A_V$ varies between 0.5 and 1.4 mag when we assume a background of 2.5 mag. There is a decreasing trend in $R_V$, from 3.7 at the tip of IC\,63 to $\sim 3.4$ when moving further away from $\gamma$~Cas. With a correction for background and/or foreground material, the latter value can go as low as $2.5$.

It is the first time that this type of measurement has been made for a PDR. This approach was possible for IC\,63 because of its low optical depth, making this PDR transparent for a sufficient number of detectable background sources. This same technique would therefore also be applicable to IC\,59. Most other PDRs, such as the Horsehead nebula, \edit2{the Orion bar, and NGC 7023,} have much higher optical depths and would not allow a sufficient amount of background stars to be measured. 
  
By combining the $A_V$ map with FIR maps from \textit{Spitzer} and
\textit{Herschel}, we were able to measure the shape of the average, $A_V$-normalized surface brightness of the dust. A dual modified blackbody model fits the shape of this curve well, and provides two temperatures (30 and 227 K) and measurements of $\kappa_{160} / \kappa_V$ (\num{6.98 \pm 0.88 e-4} and \num{2.91 \pm 0.53 e-9}).
By performing fits of the same model to individual pixels of the $A_V$ map, we derived a map of $\kappa_{160} / \kappa_V$.
The values of this map are of the same order of magnitude, but there are local variations which differ significantly (a factor 2 or 3) from the average.
\edit1{Both the $R_V$ map and the $\kappa_{160} / \kappa_V$ show variations, but no correlations were found between the two maps, and it is unclear if there are relationships with other physical quantities.}

\acknowledgments We thank the referee for a very thorough report, leading to significant improvements to the quality and completeness of this work. This research was made possible thanks to a BOF (bijzonder onderzoeksfonds) fellowship, a special research fund provided by Ghent University.

\software{DOLPHOT \citep{2000PASP..112.1383D, 2016ascl.soft08013D}.
  Tiny Tim \citep{2011SPIE.8127E..0JK}.
  BEAST \citep{2016ApJ...826..104G}.
  This research made use of Astropy,\footnote{http://www.astropy.org} a community-developed core Python package for Astronomy \citep{2013A&A...558A..33A, 2018AJ....156..123A}}

\appendix

\section{New BEAST feature: parallelization using subgrids}
By introducing the stellar distance as an extra parameter, the dimensionality of the whole SED
model grid is increased from 6D to 7D. Additionally the range of distances is potentially very
large (see Table \ref{tab:grid}), and because the distance rescales the whole SED, a granularity on par
with the $A_V$ grid is necessary. Since we chose to use $\sim 40$ (log spaced)
distance bins, the amount of memory needed to work with the grid is also increased 40-fold. Originally we would split the catalog into many pieces, and then let every process work on a
small part of the catalog. This approach is problematic because for every process launched, the
whole grid has to be loaded into memory. Therefore, the total memory of the machine still imposes a limit on the number of
fitting processes that can be run simultaneously.

To deal with these large grids, a memory-saving parallelization strategy was developed, which we
call the \textit{subgridding} approach. This approach entails that each process will work on
only a small portion of the grid or \textit{subgrid}. This works for the physics model setup,
the noise model generation, as well as the fitting step. For each subgrid, a separate set of statistics and 1D posterior distributions is obtained. By properly taking into account the weight (integrated posterior) of each subgrid and combining these individual 1D posterior distributions, the statistics can be recalculated to reflect those of the whole grid. 

To facilitate this
parallelization scheme, a set of functions for splitting and/or merging the model grids and
output statistics was developed, and an example workflow is given in the
documentation.\footnote{\url{https://beast.readthedocs.io/en/latest/subgrid_parallelism.html}}
If the size of the RAM proves problematic for running the BEAST, then the number of subgrids can simply be increased. Arguably, this makes the BEAST scalable to much larger grid sizes. 

\section{New BEAST feature: spatial variation of noise model}
Since the extended emission varies for different locations in the nebula, the noise model should
ideally depend on the position of the source that is being fit. To investigate the spatial
dependence of the noise model, a set of tools was developed and added to the BEAST. The concept
is that a number of regions on the sky that have similar intensities of background/foreground emission are
determined (background regions). Within each of these background regions, a separate
set of ASTs is performed, and an individual noise model is created. If the resulting noise
models differ significantly, an individual fitting run can be done for each background
region, using the corresponding noise model to fit only the sources that fall within that
region.

\subsection{Background map tool}
\begin{figure*}[tbh]
  \centering
  \plottwo{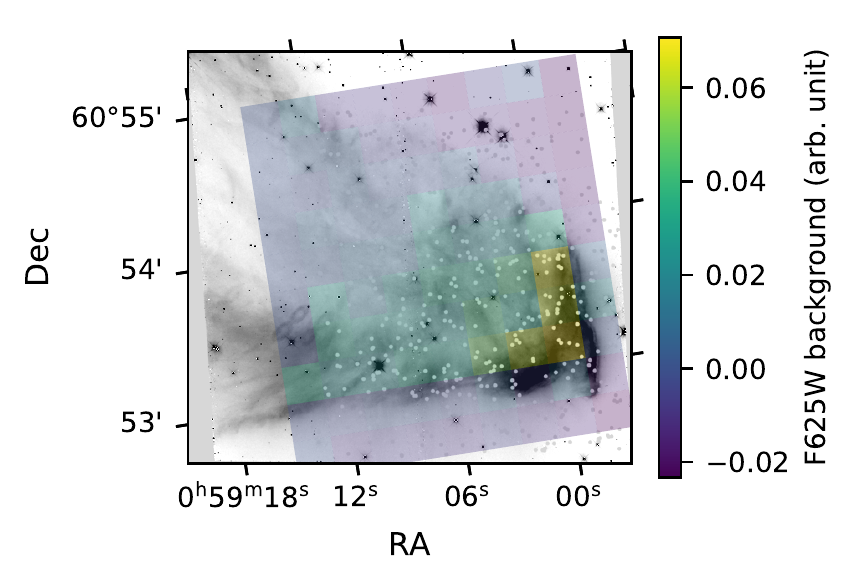}{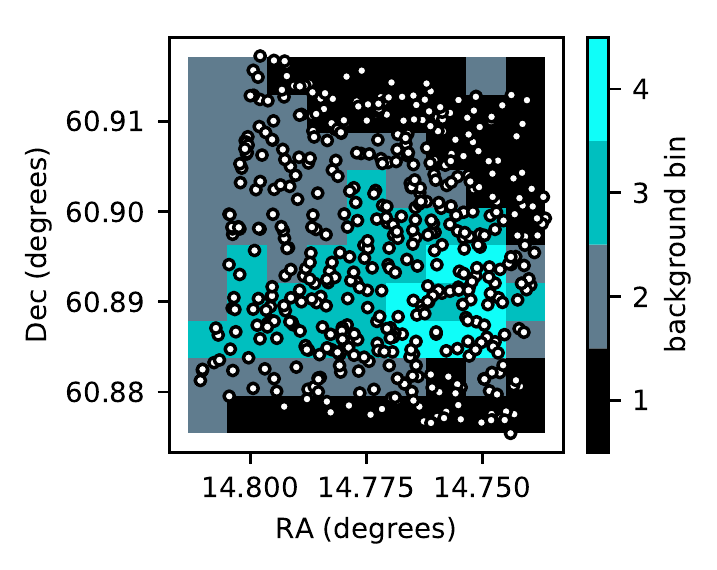}
  \caption{\textit{Left:} Illustration of the background map tool. In this case, the F625W image
    was used as an input, so the background intensity mainly consists of H$\alpha$ emission. The
    point sources have been masked using the positions listed in the cleaned up catalog
    described in section \ref{subsub:culling}. This can be seen from the empty discs on the
    image. The resolution of the map is chosen so that there's at least a handful of
    sources in each \edit1{pixel}. \textit{Right:} Example of a set of spatial regions, derived from the
    background map on the left. The colors indicate which background intensity regime each
    spatial bin and, by extension, each source (\textit{circles}) belongs to.}
  \label{fig:backgroundmap}
\end{figure*}

The first tool consists of a stand-alone script which can generate a map, with pixels of a
user-specified size along the Right Ascension (RA) and Declination (Dec) axes. To measure the
background for a single \edit1{pixel}, it is first determined which of the sources of the input catalog
fall into this bin, according to their positions. Then, a reference image is used (the F625W
image in our case) to measure the background for each source. The sources themselves are masked
out, and their background is determined using an annulus of a certain radius and thickness.
Determining an ideal inner and outer radius for the annuli may require some trial and error by
the user. The background for this \edit1{pixel} is then set to be the median of these individual
background measurements. For IC\,63, the resulting background map is shown in the left panel of
figure \ref{fig:backgroundmap}.

On a side note, this same tool can also be used to construct a map of point source densities in
the exact same format. This is useful for catalogs that have a very high but variable density of
stars, leading to different crowding effects depending on the position. The source density is
derived directly from the positions listed in a given catalog. Since the rest of the code is
agnostic of the quantity described by either map, a common approach can be used to model either
effects of extended emission or those of crowding.

\subsection{Selection of AST positions}
One of the applications that use this background map is the routine for selecting the positions
of the ASTs. We want to make sure that the areas corresponding to different levels of background
intensity are all sufficiently sampled. We achieve this by first choosing a set of bins (usually
a small number, due to the available hard disk space and computing time), each of which
represents a certain range in background intensity.
Then, the \edit1{pixels} of the map described above
can be distributed among these bins, leading to groups of \edit1{pixels} which constitute the desired
background regions (Figure \ref{fig:backgroundmap}). A fixed set of SEDs is reused for
each region, and each SED is assigned a random position within that region. This way, each
background region has the same number of samples regardless of its total surface area.

\subsection{Creating individual noise models} Once the selected ASTs have been run through our photometry routine, and a
catalog of fake stars has been produced, both the observed and the fake catalog are split up
according to the background regions (see again figure
\ref{fig:backgroundmap}, right panel). By applying the same spatial split to both the observed
catalog and the fake catalog, a series of catalog pairs is created. The BEAST can
then use each pair consisting of one fake and one observed catalog to generate the individual noise models, and perform the individual fitting
runs if the resulting noise models are found to be significantly different.

\begin{figure*}
  \centering
  \plotone{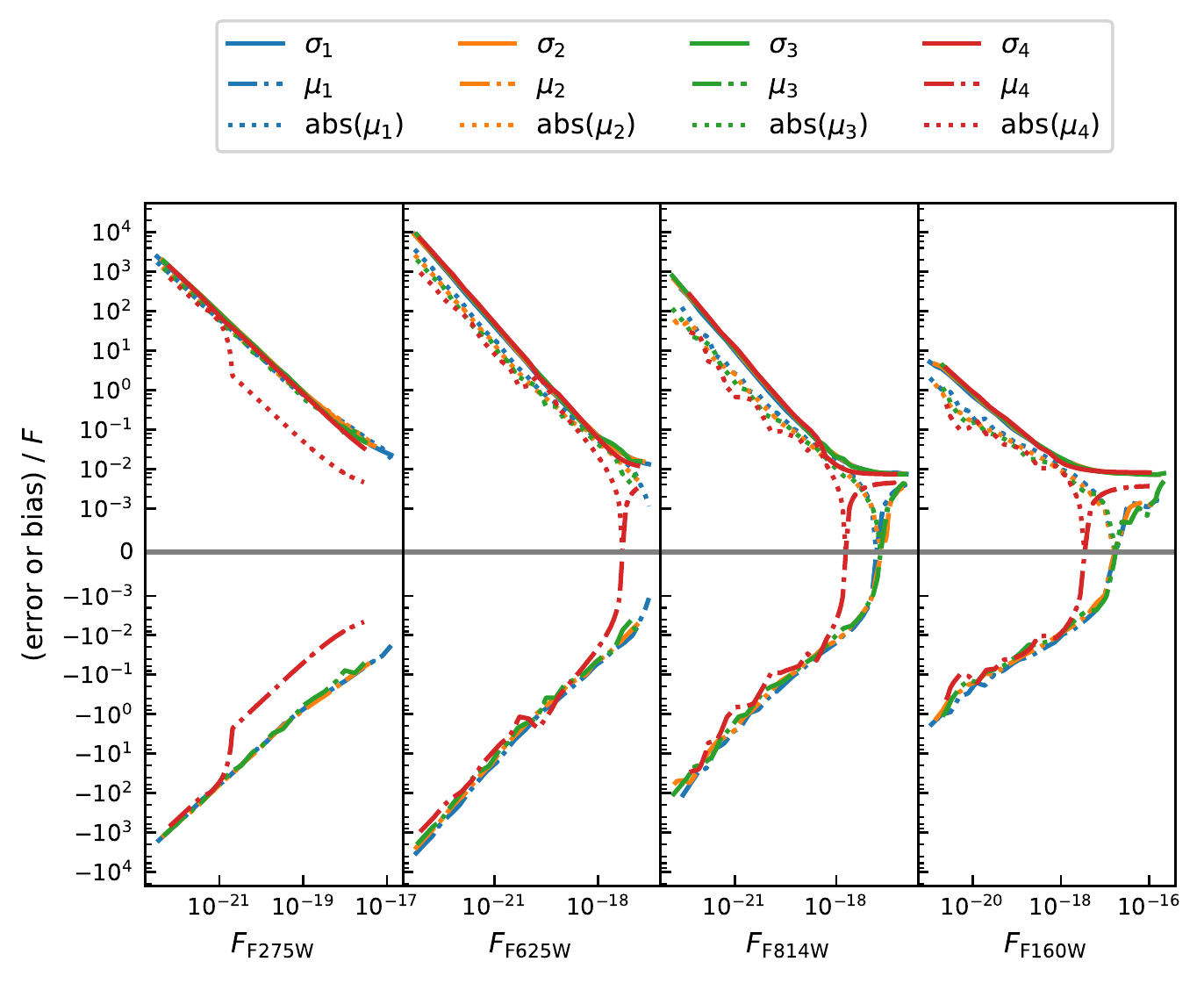}
  \caption{Four different noise models, one for each background region. The numbers 1,2,3,4 point to the regions shown in the right panel of
    Figure \ref{fig:backgroundmap}.}
\label{fig:extranoise}
\end{figure*}

Using the regions shown in the right panel of Figure \ref{fig:backgroundmap}, we created four noise
models (Figure \ref{fig:extranoise}). Despite the differences between the noise parameters for the background bins, we found that the predicted $A_V$ and $R_V$ are not significantly influenced by using these extra noise models. In each bin there are at most four stars which have an $A_V$ or $R_V$ that differs more than 10\%. For some of the other parameters the differences can be larger, but the fit quality is not necessarily better. Therefore, the results presented in Section \ref{sec:results} and the rest of the analysis are based on fits using the single noise model that was shown in Figure \ref{fig:toothpick}. Note that this noise model still benefits from the background-based AST position picking, because it makes sure that the tip of the nebula is sufficiently sampled.

\section{The $A_V$ and $R_V$ maps in table format}
We provide the values and
standard deviation for $A_V$ and $R_V$ of the maps in Tables \ref{tab:avnumbers} and \ref{tab:rvnumbers}.
The RA and Dec values in the table are the coordinates of the bottom left corner of each pixel, with the tables oriented in
the same way as the color maps in Figure \ref{fig:maps}.

\begin{deluxetable*}{c|cccccc}
  \tablecaption{Av map and uncertainty (same orientation as Figure \ref{fig:maps}) \label{tab:avnumbers}}
  \label{tab:avnumbers}
\tablehead{\colhead{Dec / RA (degrees)} & \colhead{14.8021} & \colhead{14.7878} & \colhead{14.7736} & \colhead{14.7593} & \colhead{14.7450} & \colhead{14.7308}}
\startdata
60.9206 & 3.54 $\pm$ ? & X & X & X & X & X\\
60.9137 & 3.32 $\pm$ 0.48 & 3.00 $\pm$ 0.22 & 3.00 $\pm$ 0.33 & 3.26 $\pm$ 0.36 & 4.75 $\pm$ 0.36 & X\\
60.9067 & 3.92 $\pm$ 0.29 & 3.10 $\pm$ 0.28 & 3.03 $\pm$ 0.25 & 3.01 $\pm$ 0.16 & 5.12 $\pm$ 0.45 & X\\
60.8998 & 3.21 $\pm$ 0.24 & 3.48 $\pm$ 0.24 & 3.60 $\pm$ 0.22 & 3.19 $\pm$ 0.19 & 2.94 $\pm$ 0.21 & 3.15 $\pm$ 0.84\\
60.8928 & 3.04 $\pm$ 0.30 & 3.28 $\pm$ 0.15 & 3.06 $\pm$ 0.19 & 3.92 $\pm$ 0.23 & 3.04 $\pm$ 0.23 & X\\
60.8859 & 2.98 $\pm$ 0.22 & 3.04 $\pm$ 0.16 & 3.30 $\pm$ 0.16 & 3.44 $\pm$ 0.36 & 3.56 $\pm$ 0.36 & X\\
60.8790 & 2.67 $\pm$ 0.12 & 2.78 $\pm$ 0.21 & 2.38 $\pm$ 0.18 & 2.29 $\pm$ 0.14 & 2.49 $\pm$ 0.15 & X\\
\enddata
\end{deluxetable*}

\begin{deluxetable*}{c|cccccc}
  \tablecaption{Rv map and uncertainty (same orientation as Figure \ref{fig:maps}) \label{tab:rvnumbers}}
  \label{tab:rvnumbers}
\tablehead{\colhead{Dec / RA (degrees)} & \colhead{14.8021} & \colhead{14.7878} & \colhead{14.7736} & \colhead{14.7593} & \colhead{14.7450} & \colhead{14.7308}}
\startdata
60.9206 & 2.76 $\pm$ ? & X & X & X & X & X\\
60.9137 & 3.44 $\pm$ 0.11 & 3.73 $\pm$ 0.14 & 3.48 $\pm$ 0.10 & 3.47 $\pm$ 0.15 & 3.75 $\pm$ 0.36 & X\\
60.9067 & 3.51 $\pm$ 0.16 & 3.34 $\pm$ 0.10 & 3.34 $\pm$ 0.10 & 3.68 $\pm$ 0.11 & 3.68 $\pm$ 0.13 & X\\
60.8998 & 3.43 $\pm$ 0.14 & 3.52 $\pm$ 0.11 & 3.45 $\pm$ 0.14 & 3.55 $\pm$ 0.07 & 3.59 $\pm$ 0.11 & 3.60 $\pm$ 0.25\\
60.8928 & 3.50 $\pm$ 0.31 & 3.58 $\pm$ 0.10 & 3.40 $\pm$ 0.09 & 3.58 $\pm$ 0.10 & 3.65 $\pm$ 0.15 & X\\
60.8859 & 3.31 $\pm$ 0.09 & 3.53 $\pm$ 0.16 & 3.72 $\pm$ 0.14 & 3.63 $\pm$ 0.15 & 3.66 $\pm$ 0.26 & X\\
60.8790 & 3.90 $\pm$ 0.06 & 3.61 $\pm$ 0.17 & 3.49 $\pm$ 0.15 & 3.61 $\pm$ 0.09 & 3.57 $\pm$ 0.11 & X\\
\enddata
\end{deluxetable*}

\end{document}